\documentclass{aastex61}
\usepackage{amsmath}
\usepackage{bm}	

\shorttitle{A catalogue of close encounter pairs}
\shortauthors{Feng \&  Jones}

\begin{document}
\title{A catalogue of close encounter pairs}
\correspondingauthor{Fabo Feng}
\email{fengfabo@gmail.com, f.feng@herts.ac.uk}
\author{Fabo Feng}
\affiliation{Centre for Astrophysics Research, School of Physics, Astronomy and Mathematics, University of Hertfordshire, College Lane, Hatfield AL10 9AB, UK} 

\author{Hugh R. A. Jones}
\affiliation{Centre for Astrophysics Research, School of Physics, Astronomy and Mathematics, University of Hertfordshire, College Lane, Hatfield AL10 9AB, UK} 
\author{Tabassum S. Tanvir}
\affiliation{Centre for Astrophysics Research, School of Physics, Astronomy and Mathematics, University of Hertfordshire, College Lane, Hatfield AL10 9AB, UK} 

\begin{abstract}
  We provide a catalogue of pairs of stars whose periapses are less than 1\,pc within the past or future 100\,Myr. We use astrometric data from TGAS, Hipparcos and radial velocity data from RAVE and XHIP to find the space motions and hence the initial conditions of 229, 441 stars in Galactic coordinates. We simulate the orbits of these stars and focus on the time, distance and velocity at periastron for 8,149 pairs using the k-d tree algorithm to find nearest neighbors. We find an anisotropy in the directions of encounter pairs caused by the solar apex motion, indicating a role of peculiar motion imposing an anisotropic tidal force on planetary systems. We call this effect ``kinematic tide''.  Among the encounter pairs there are 4 encounters with the Solar System with periapses less than 1 pc and 96 pairs with periapses less than 0.1 pc. We also find 577 close encounters of stars which host planetary systems and/or debris disks. We discuss a range of uses for an encounter catalogue and present an example of how the time-varying network of stellar motions will be crucial for efficient interstellar travel between planetary systems. The catalogues are available at http://star.herts.ac.uk/pandora/cep1.
\end{abstract}
\section{Introduction}     \label{sec:introduction}

The habitability and dynamics of the Solar System are influenced by close encounters during the Sun's migration in the Galaxy. The encounters can strongly perturb the Oort cloud, deliver long period comets in Earth-crossing orbits and bombard the Earth, leading to catastrophic events such as mass extinction \citep{feng13}. Encounters may also be responsible for the formation of Sedna-like objects which become detached from Neptune and are immune from the tidal force from the Galaxy \citep{jilkova15}. Although planet nine has been proposed to explain the orbital alignment of Sedna-like objects \citep{batygin16}, planet nine itself was probably captured from a close encounter \citep{mustill16}. The habitability and dynamics of any planets belonging to stars involved in close encounters, especially small ones, would also be influenced by the Solar System. The tidal force from the Sun could strongly perturb any disc material in the encounter systems, increase the impact rate, and even alter the orbits of outer planets orbiting low-mass stars. Hence the discovery of close encounters of the Solar System and further studies of the dynamical interaction between encounter stars and the Solar System will advance our understanding of the habitability and dynamical history of the Solar System and its past or future neighborhood.

The frequency of close encounters may be high over the past or future few million years based on the discoveries of close encounters such as Scholz's star and GJ 710, which passed and will pass the Sun at respective distances of 52\,kAU and 13\,kAU, 70,000 yr ago and 1.35 Myr from the present \citep{mamajek15,berski16}. In particular, GJ 710 will penetrate and strongly perturb the Oort cloud leading to at least an order of magnitude increase in the flux of new long-period comets. Any planetary system belonging to GJ 710 would also be significantly influenced by the comet shower from the Oort cloud and as would any similar structures existing in the GJ 710 system.

Anthropically, we have first been concerned about stellar encounters with the Solar System, however, such near encounters of stars and their planetary systems in the past and future maybe considered as ubiquitous in our dynamically active Galaxy. Although field stars typically have a relaxation time longer than the age of the Galaxy \citep{binney08_book}, the cumulative and stochastic effects of stellar encounters could have significant influence on the outer part of a stellar system or on multiple-component stellar systems. With more and more findings of widely separated stellar or planetary companions (e.g. \citealt{gizis01}, \citealt{dhital11} and \citealt{andrews17}), it can be expected that the interaction with field encounters will be critical for some systems (e.g. Proxima Centauri, \citealt{feng18}).

Although the field encounters of a single star are still too distant to have strong dynamical impact on the stellar system, they may influence the habitability of the planets in the system through perturbing the potential circumstellar disc (or Oort cloud) and thus triggering asteroid or comet showers \citep{heisler87}. Stellar encounters and the Galactic tide could also perturb planets on wide orbits and vary their periapses and eccentricities, leading to a variation of the strength of perturbations on inner planets from outer planets, and thus change the motions and habitability of inner planets \citep{kaib13}. The planetary habitability could also be influenced by supernovae through direct damage of organisms or tropospheric ionization caused by cosmic rays from a supernova remnant if a planetary system is less than 100\,pc from the progenitor \citep{thomas16}. Other interactions such as encounters with globular clusters or molecular clouds could also significantly influence planetary habitability (e.g. \citealt{feng13} and \citealt{domainko13} and see \citealt{bailer-jones09} for a review). 

Previous studies of the interaction between stars (or ``encounter pairs'') are typically limited to the statistical property of a system. For example, the influence of stellar encounters on planet formation in star clusters has been studied using N-body simulations \citep{malmberg07}. The study of individual encounters has only been done for the Solar System in order to explore the interstellar effects on the Earth's habitability \citep{feng15b}. These limitations in previous studies are partly caused by the assumption that field encounters of a star are dynamically weak and thus cannot significantly affect its evolution. However, with increasing number and diversity of the planetary components of stellar systems and of the habitability of exoplanets, the influence of encounters becomes an interesting consideration.

Encounter pairs also provide a time-varying network of potential connections between stars. In other words, the neighborhood of a star is dynamic rather than static. Although Proxima is our current closest neighbor, GJ 710 will be 20 times closer than Proxima about 1.35\,Myr from now. If planets around encounter pairs are found to be habitable, they would form a time-varying network of connections between habitable worlds. Studies of this network are necessary to explore optimal travel routes for efficient interstellar exploration, which becomes interesting, for example, due to the practical endeavor of the Breakthrough Starshot project - \url{https://breakthroughinitiatives.org/} and academic research (e.g., \citealt{heller17}). Moreover, encounters potentially play a role in the transport of life to the Earth or the other way around \citep{napier04,wickramashinghe10, lingam17}.

In recent years there have been huge improvements in the available astrometric and radial velocity data in the Galactic field particularly courtesy of Gaia TGAS \citep{gaia16}, RAVE DR5 \citep{kunder16} and LAMOST DR2 \citep{luo16} as well as other catalogues. This enables us to compute the orbits of more than 0.2 million stars to find potential encounter pairs separated by less than one parsec and for which the uncertainty of periastron can be estimated through direct simulations of 1000 clones for each encounter pair. Thus we are able to construct a large scale catalogue of encounter pairs, we call this catalogue CEP1 (Close Encounter Pairs 1). Although there are wide range of applications for such a catalogue, we select encounters of the Sun as well as stars hosting exoplanetary systems as being of particular interest. This work provides a prototype for more advanced encounter catalogs based on the upcoming Gaia and ground-based spectroscopy data releases (e.g. FunnelWeb,  \citealt{lawson16}). 

This paper is structured as follows. In section \ref{sec:data}, we introduce the catalogues we use and combine to form the catalogue of 0.24 million stars. In section \ref{sec:method}, we introduce the Galactic model, numerical methods and Monte Carlo approach, which are used for the integration of stellar orbits. In section \ref{sec:results}, the catalogue based on the numerical simulations is described and individual encounter pairs are discussed in detail. In section \ref{sec:application},  we discuss the range of potential applications of our encounter pairs catalogue. Finally, section \ref{sec:conclusion} presents our conclusions.
\section{Data}\label{sec:data}
To form a catalogue of stars with position and velocity information, we crossmatch the catalogues of the Tycho-Gaia Astrometric Solution (TGAS, \citep{gaia16}), Radial Velocity Experiment fifth data release (RAVE5, \citealt{kunder16}), new Hipparcos astrometric catalogue (HIP2; \citealt{leeuwen07}), extended Hipparcos compilation (XHIP; \citealt{anderson11}), Large Sky Area Multi- object Fiber Spectroscopic Telescope second data release (LAMOST2; \citealt{luo16}), URAT Parallax Catalogue (UPC; \citealt{finch16}), parallax data from REsearch Consortium On Nearby Stars (RECONS; \citealt{winters15}), proper motions from the fifth US Naval Observatory CCD Astrograph Catalogue (UCAC5; \citealt{zacharias17}), Hot Stuff for One Year (HSOY; \citealt{altmann17}), PPMXL \citep{roeser10}, and USNO-B1.0 \citep{monet03} as follows.
\begin{itemize}
\item  RAVE5+TGAS (210,099 targets; denoted by {\it rt}) - The common objects shared by RAVE5 and TGAS, with radial velocities from the former and astrometric data from the latter.
\item XHIP+TGAS (30,849 targets; denoted by {\it xt}) - The astrometric parameters in the XHIP catalogue are replaced by the TGAS ones for common objects.
\item  LAMOST2+TGAS (87,781 targets; denoted by {\it lt}) - TGAS is cross-matched with LAMOST2 with a search-cone of 5$''$\footnote{This criterion is relatively conservative and is able to exclude many nearby wide binaries and open clusters, which are false positives of encounter pairs.}.
\item RAVE5+HIP2 (5387 targets; denoted by rh). RAVE5 and HIP2 are cross-matched with a cone of 5$''$.
\item  XHIP+HIP2 (15,079 targets; denoted by {\it xh}) - Instead of using the proper motions in Tycho-2 \citep{hog00}, we consistently use HIP2 proper motions for targets in XHIP because it contains covariance matrix of astrometric parameters, following \cite{dybczynski15}. 
\item LAMOST2+UPC (4242 targets; denoted by {\it lu}) - LAMOST2 and UPC are cross-matched with a cone of 5$''$.
\item RAVE5+RECONS+HSOY+UCAC5+PPMXL+USNO-B1.0 (53 targets; denoted by {\it rr}) - RAVE5 and RECONS are cross-matched with a cone of 5$''$. This catalogue is further cross-matched with HSOY, UCAC5, PPMXL and USNO-B1.0 with a 5$''$ cone to obtain precise proper motions.
\end{itemize}
  
The above catalogues are sorted in priority order. For example, if the same target appears in {\it rt} and {\it xt}, the parameters in {\it rt} are used. Therefore the combination of the above catalogs forms a catalogue of 6D astrometric parameters of 349,161 stars. The positions of stars in each catalogue are converted to epoch J2000 in the ICRS coordinate system. To form a more conservative catalogue and thus make the identification of encounter pairs more efficient, we only simulate the motions of targets with relative proper motion and parallax errors less than 100\%, with radial velocity (RV) less than 100\,km/s, with RV error less than 10\,km/s, and with good RV quality\footnote{We select XHIP targets with RV quality denoted by ``A''  and choose RAVE targets with internal errors less than 10\,km/s. }, leading to a sample of 229, 441 stars. It should be noted that the periastrons of encounter pairs are very sensitive to observational uncertainties. For example, if the radial velocity and parallax are relatively certain while the proper motions are measured with a relative error of 100\%, the error of perihelion for a encounter of the Solar System would also be $\sim$100\%. When considering errors on parallaxes and radial velocities, the uncertainty of perihelion becomes even more significant. However, previous studies suffer from a lack of good quality data and typically use all available data to search for encounters of the Solar System. This may lead to many false positives, which are rejected by later studies (e.g. \citealt{crifo17}). Since most stars in our catalogue are bright and are in the solar neighborhood, we call this catalogue Fellow Stars with 6D astrometric parameters (dubbed FS).

The TGAS catalogue provides the covariance matrix for the five astrometric parameters, $\alpha$ (right ascension), $\delta$ (declination), $\pi$ (parallax), $\mu_{\alpha}$ (proper motion in the right ascension), and $\mu_{\delta}$ (proper motion in the declination). For stars in the HIP2 catalog, we derive the covariance matrix using the method introduced in Appendix B of \cite{michalik14}. We further assume that the five astrometric parameters are independent with the radial velocity, $v_r$ . For stars without covariance information, we assume independent errors for all six parameters.

\begin{figure}
  \centering
  \includegraphics[scale=0.5]{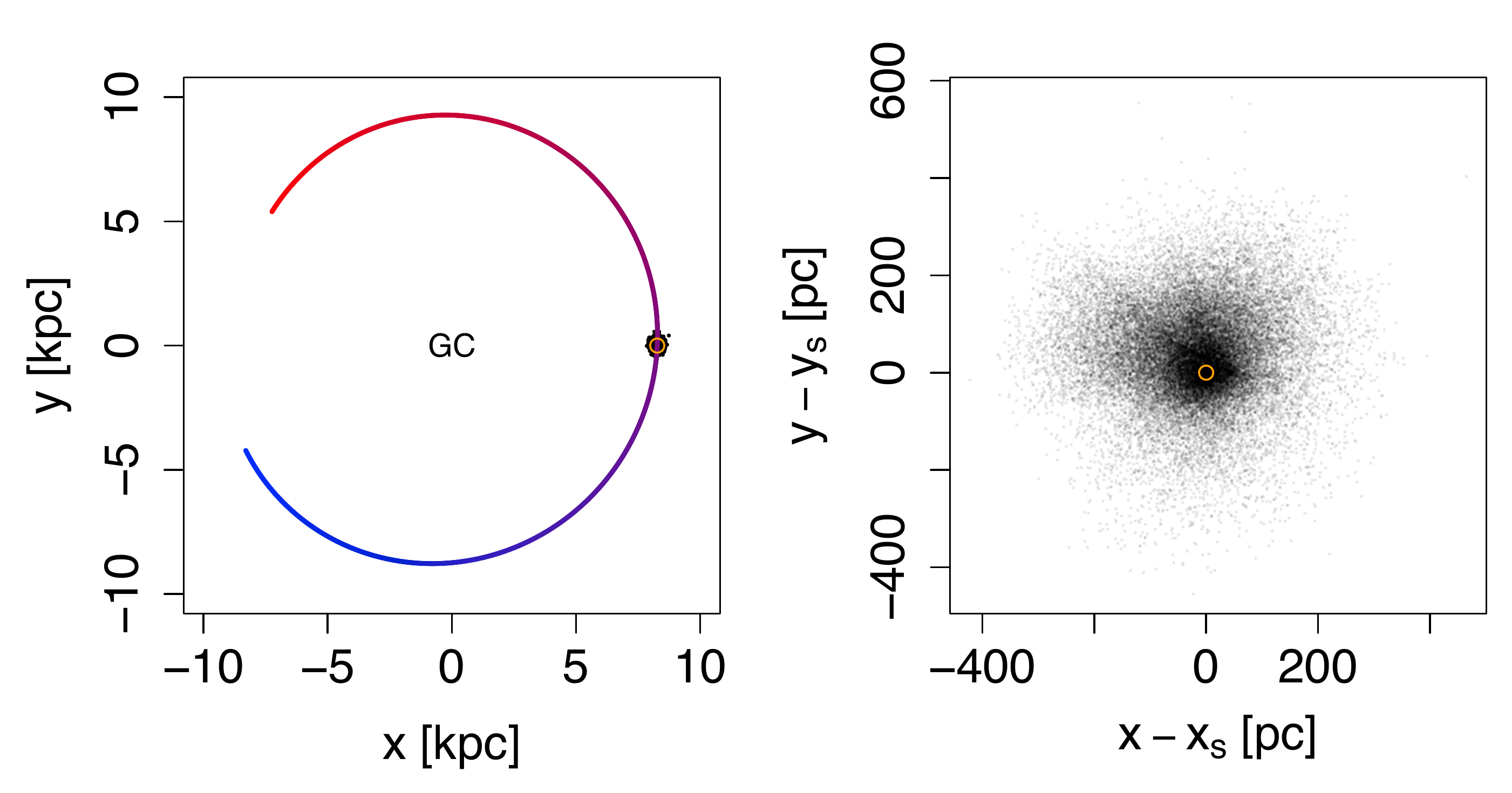}
  \caption{Left panel: orbit of the Sun from -100\,Myr to +100 Myr and FS stars with relative astrometric errors less than 10\%. The orange circle denotes the current location of the Sun. Right panel: The spatial distribution of these FS stars in the heliocentric reference frame in the Galactic plane.}
  \label{fig:solar_orbit}
\end{figure}

Then we convert the 6D astrometric parameters into the initial conditions of Galactocentric (GC) positions and velocities, $x$, $y$, $z$, $v_x$, $v_y$ , and $v_z$ . The coordinate system is shown in Fig. \ref{fig:solar_orbit}. In the coordinate transformation, we adopt the 6D initial condition of the Sun according to \cite{dehnen98b,majaess09,schoenrich10,schoenrich12} and shown in table 3 of \cite{feng14}. In Fig. \ref{fig:solar_orbit}, we see that most stars with small astrometric errors are located in the solar neighborhood which is unsurprising since they need to be bright enough for precise astrometric measurements by optical surveys.

\begin{figure}
  \centering
  \includegraphics[scale=0.6]{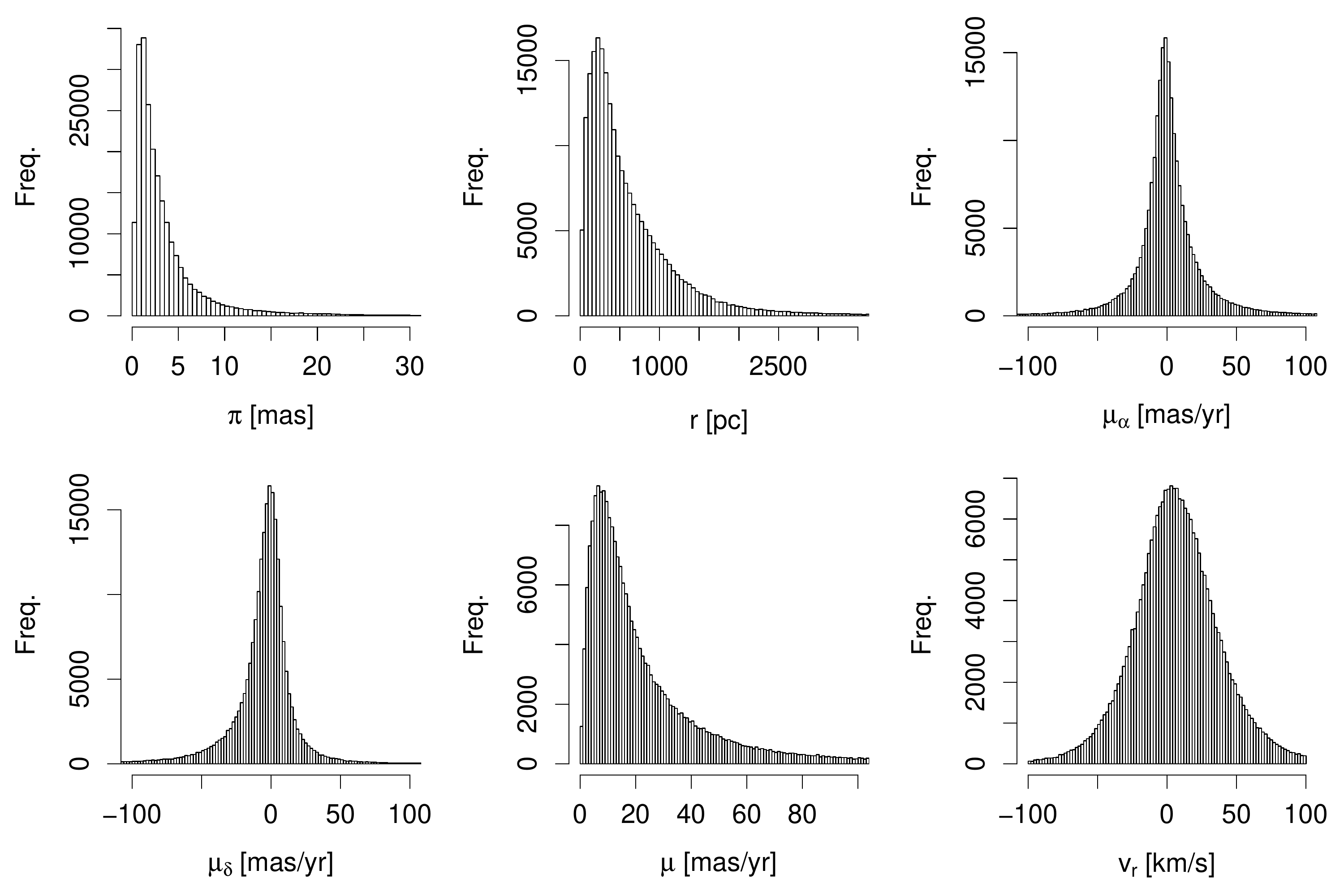}
  \caption{Distributions of FS stars over $\pi$, $r\equiv1/\pi$, $\mu_{\alpha}$, $\mu_\delta$, $\mu\equiv\sqrt{\mu_{\alpha^*}^2+\mu_\delta^2}$ and $v_r$. }
  \label{fig:astrometry_dist}
\end{figure}

We investigate the distribution of the proper motions, parallaxes, and radial velocities of the FS sample and show them in Fig. \ref{fig:astrometry_dist}. We see the concentration of targets within 300\,pc, indicating significant incompleteness of stars with distance larger than 300\,pc. We find a slight deviation from an isotropic distribution of tangential motions since the mean value of $\mu_{\alpha}$ and $\mu_{\delta}$ are 0.00 and -8.55\,mas/yr, respectively. We also find a positive mean radial velocity of 4.17\,km/s of the sample.  These anisotropies are probably due to the peculiar motion of the Sun. Assuming an isotropic spatial and velocity distribution of stars in the local standard of rest (LSR), the motions of stars in the helio-static framework would not be isotropic due to a subtraction of the solar apex motion from the velocities with respect to LSR. But this effect is reduced by the observation limit in various surveys. Since these limitations are typically caused by the distance and kinematics of stars with respect to the Sun, the corresponding catalogue tends to include stars with kinematics, which are isotropic in the helio-static reference frame. We will further discuss the influence of solar apex motion in section \ref{sec:anisotropy}. 

\section{Method}\label{sec:method}
To integrate stellar orbits numerically, we adopt an axisymmetric Galactic potential composed of budge, disc and halo defined in \cite{feng14} with corresponding parameters from \cite{sanchez01}. Then we numerically integrate the nominal orbits of FS stars using the {\small Adams} method implemented in the {\small R} package {\small deSolve}. We simulate the motions of stars from -100\,Myr to +100\,Myr with a time step of 1 kyr, leading to a relative numerical error of angular momentum of about $10^{-7}$ over a 100 Myr integration. We then find the nearest neighbors within one parsec from each star for each time step using the {\small RANN} package in {\small R}, which is an implementation of the {\small kd-tree} algorithm. Finally, we compare the distances of nearest neighbors for all time steps to select the closest stellar encounters of each star.

Considering that the 1\,kyr time step may not properly resolve the periastron, we refine the calculation of periastron using the linear approximation at the point closest to periastron. The refined time, $t_{\rm enc}$, and distance, $d_{\rm enc}$, at the periastron are
\begin{eqnarray}
  t_{\rm enc}&=&-\dfrac{{\bm r_0}{\bm v_0}}{v^2_0}~,\nonumber\\
  {\bm d}_{\rm enc}&=&{\bm r_0}+{\bm v_0}t_{\rm enc}~,\nonumber\\
  {\bm v}_{\rm enc}&=&{\bm v_0}~,
 \label{eq:linear}
\end{eqnarray}
where ${\bm r_0}$ and ${\bm v_0}$ are respectively the position and velocity of the encounter with respect to the target star. Our approach is similar to the heliocentric linear approximation used by \cite{bailer-jones15} to refine perihelia of encounters.

We estimate the uncertainty of periastron for each encounter pair in a Monte Carlo fashion. To make the estimation computationally efficient, we generate 1000 clones of each encounter pair, and numerically integrate their orbits with a time step of 0.1\,Myr (i.e. $dt=0.1$\,Myr). In addition to the 1000 clones, we also integrate the nominal orbit with the same time step. We shift the periastron of the nominal orbit and clones such that the nominal periastron computed with $dt=0.1$\,Myr is the same as the one computed with $dt=0.001$\,Myr. We have also tried smaller time steps to calculate the uncertainty of periastron, the difference is not significant. 

Since the astrometric errors are large for some distant and faint stars, the nominal periastron, $d_{\rm enc}^{\rm nom}$, may not reside within the interval defined by the 5\% and 95\% quantiles. This is partly caused by the inefficient sampling of the periastron distribution especially if the nominal periastron is far less than 1\,pc. To account for this, we set the 5\% quantile to be 0 if it is larger than the nominal periastron. We also follow \cite{dybczynski15} to estimate the mean $d_{\rm enc}$ using
\begin{equation}
  \bar{d}_{\rm enc}=\sqrt{\bar{x}^2_{\rm enc}+\bar{y}^2_{\rm enc}+\bar{z}^2_{\rm enc}}~,
  \label{eqn:mean}
\end{equation}
where $\bar{x}$, $\bar{y}$, and $\bar{z}$ are the mean coordinates of periastrons. 

\section{Results}\label{sec:results}
We apply the method described in section \ref{sec:method} to the data introduced in section \ref{sec:data} to produce the so-called ``EP1'' catalog. This catalogue contains 186,423 encounter pairs, including 48 encounters approaching the Sun with $d_{\rm enc}<1$\,pc and 24 pairs separated by less than 0.01\,pc. Since the uncertainty of encounter parameters for most stars in this catalog is too large, we make a more conservative and reliable catalog called ``CEP1'' by selecting pairs with RV errors less than 10\,km/s, and excluding UPC targets, most of which are found to have unreliable parallaxes. We further remove pairs with the mean and 90\% quantile of $d_{\rm enc}$ less than 2\,pc and 10\,pc, respectively. The CEP1 catalog contains 8149 pairs, including 4 encounters of the Sun, 96 encounters with periastron less than 0.1\,pc and 577 encounters of stars which host planets and/or debris disks.  The EP1 and CEP1 catalogs are available at \url{http://star.herts.ac.uk/pandora/cep1}. Hereafter we will analyze the CEP1 catalog.

\begin{figure}
  \centering
  \includegraphics[scale=0.6]{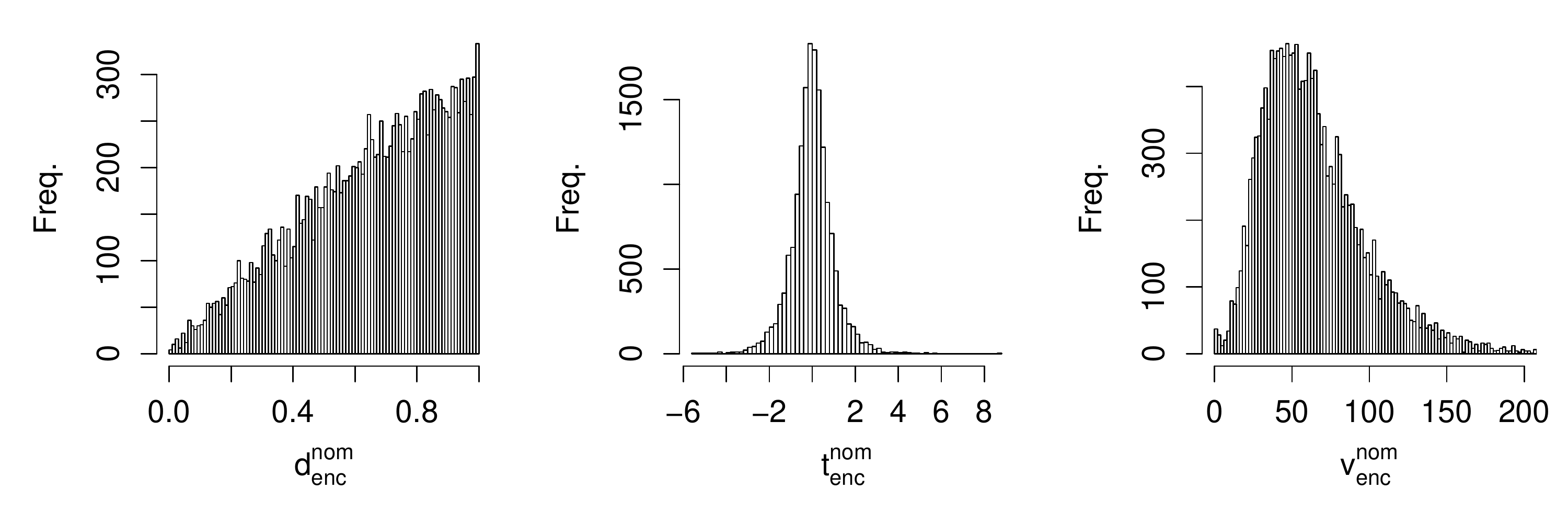}
  \caption{Distributions of encounter pairs over $d_{\rm enc}^{\rm nom}$ (left), $t_{\rm enc}^{\rm nom}$ (middle), and $v_{\rm enc}^{\rm nom}$ (right). Each encounter pair is counted twice due to a consideration of the order of pair components.}
  \label{fig:hist_dtv}
\end{figure}

In the CEP1 catalog, we find the number of siblings of a target based on a cross-match between CEP1 and the Simbad database provided by the CDS service\footnote{\url{http://simbad.u-strasbg.fr/simbad/}}. By excluding targets with multiple siblings (e.g., cluster stars and binaries) from the CEP1 catalog, we show the distribution of various quantities for single stars in Fig. \ref{fig:hist_dtv}. The distribution of periastron $P(d_{\rm enc}^{\rm nom})$ is found to be proportional to $d_{\rm enc}^{\rm nom}$, as shown in the left panel of Fig. \ref{fig:hist_dtv}. This is expected because $v_{\rm enc}$ is independent with $d_{\rm enc}$ and $P(d_{\rm max})\propto n \bar{v}_{\rm enc}d_{\rm max}$ \citep{feng14}, where $\bar{v}_{\rm enc}$ is the mean velocity of an encounter with respect to the target star, $d_{\rm max}$ is the maximum periastron, and $n$ is the local stellar number density. In Fig. \ref{fig:hist_dtv}, we also observe that $t_{\rm enc}^{\rm nom}$ is centered around 0, indicating significant incompleteness for encounters with $|t_{\rm enc}|>1$\,Myr. Moreover, we find a mean encounter velocity, $\bar{v}_{\rm enc}$, of 65\,km/s which is larger than the mean velocity dispersion (about 53\,km/s) of stars in the solar neighborhood \citep{rickman08}. This is partly caused by the fact that encounters tend to approach a star from the opposite direction to its peculiar velocity, $\bm v_p$. Another reason is that fast moving stars pass other stars more frequently, leading to a shift of $P(v_{\rm enc})$ to a higher $v_{\rm enc}$. 

\subsection{Anisotropic encounters}\label{sec:anisotropy}
Isotropic encounter velocities are typically assumed in the studies of perturbations of the Oort cloud, dynamical evolution of globular clusters, and the relaxation of stellar systems in general. However, encounter velocities are not isotropic due to the peculiar motion of the reference object with respect to the LSR \citep{feng14}. We call this effect ``kinematic tide''. This effect is different from the dynamical friction \citep{chandrasekhar43}, although both of them are caused by the motion of a reference object and both lead to anisotropic perturbations on themselves.

\begin{figure}
  \centering
  \includegraphics[scale=0.5]{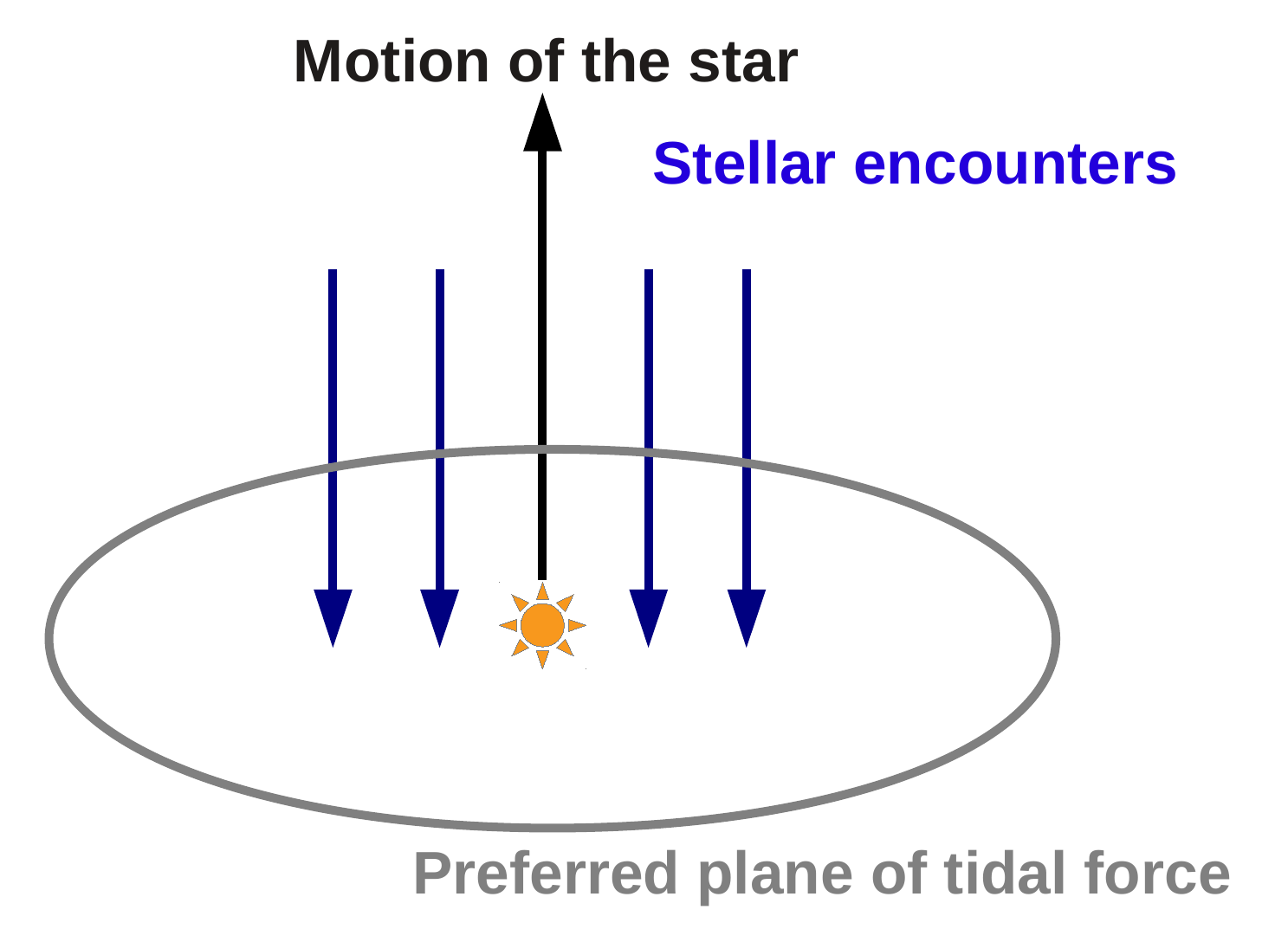}
  \caption{Tidal plane induced by the peculiar motion of a star.}
  \label{fig:tidal_plane}
  \end{figure}

Dynamical friction is caused by the anisotropic perturbations on the reference object from anisotropic spatial distribution of encounters. Relatively speaking, a massive object attracts encounters during its motion, leading to an over-dense region behind it and an under-dense region before it, leading to an anisotropic gravitational force on the reference object and thus a deceleration of the object. Since the anisotropy is correlated with the mass of the reference object, dynamical friction is only important for massive objects. Unlike dynamical friction, kinematic tide is caused by kinematics of stars. The anisotropy in the distribution of encounter periapses is more significant for faster stars. For example, if all stars are static apart from a reference star, the reference star would only be encountered by stars from the opposite direction of the reference star's motion. Thus the periastron would only be in the plane that perpendicular to the star's motion. We call this plane ``tidal plane'', which is induced by the peculiar motion of a star and is illustrated in Fig. \ref{fig:tidal_plane}.

\begin{figure}
  \centering
  \includegraphics[scale=0.6]{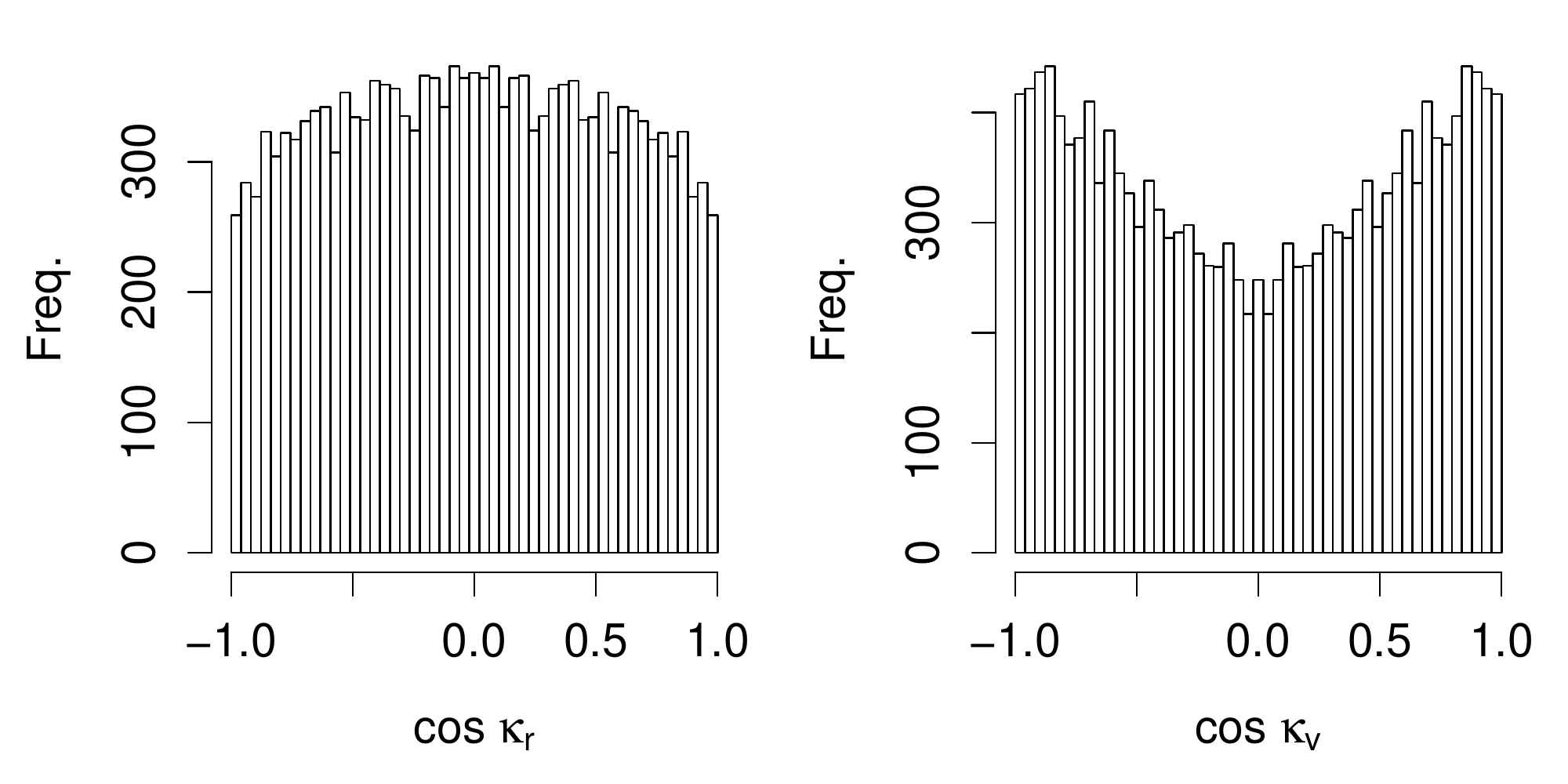}
  \caption{Distributions of the directions of periastron $\kappa_r$ (left) and relative velocity $\kappa_v$ (right) of encounter pairs with respect to the solar apex.}
  \label{fig:anisotropy}
\end{figure}

This anisotropy in the distribution of periastrons is well illustrated by the distribution of the directions of periastrons and relative velocities of encounter pairs with respect to the solar apex determined by the peculiar velocity of the Sun \citep{schoenrich10}, shown in Fig. \ref{fig:anisotropy}. If the encounters were isotropic, $P(\cos{\kappa_{\rm r}})$ and $P(\cos{\kappa_{\rm v}})$ would be uniform. But the non-uniformity in the distributions is significant. Although the motions of nearby stars are approximately isotropic in the LSR reference frame, the solar apex motion is added onto the peculiar motions of stars to derive initial conditions for simulations. Hence the solar apex motion is manifested in the distribution of the periastrons and relative velocities of encounter pairs. The non-uniform angular distribution of periapses shown in Fig. \ref{fig:anisotropy} is similar to the angular distribution of perihelia shown in fig. 7 of \cite{feng14} based on simulated encounters. 

The effect of stellar peculiar motion could be significant. For example, the planetary system would be preferentially perturbed by encounters around the tidal plane. This effect is similar to the Lidov-Kozai mechanism \citep{lidov62,kozai62}, which is a secular perturbation on a satellite from a third body on a wide orbit. However, this effect is stochastic while the Kozai mechanism is deterministic. Moreover, the tidal force is axisymmetric on average in the tidal plane while the tidal force from a third body is periodic and is not axisymmetric even averaged over several orbits due to the eccentricity of the third body. In summary, studies of the influence of kinematic tide on the formation and evolution of stellar systems are warranted in the case of encounter pairs especially where they harbour discs and planetary systems.

\subsection{Encounters of the Solar System}\label{sec:solar}
As an example, the four close encounters of the Solar System with $d_{\rm enc}<1$\,pc are shown in Table \ref{tab:sample}. Since we choose stars with small measurement errors, we only find four encounters with nominal perihelion less than 1\,pc. In these four encounters, HIP 71681 and HIP 71683 are bound to form the Alpha Centauri system. If we calculate the motion of the barycenter of Alpha Centauri A and B with respect to the Sun, we find $d_{\rm enc}^{\rm nom}=0.97$\,pc and $t_{\rm enc}^{\rm nom}=0.03$\,Myr. Because the barycentric velocities of alpha Centauri A and B are much less than the heliocentric motion of their barycenter, the values for HIP 71681 and 71683 reported in \ref{tab:sample} are reliable. This principle also appliable for other wide binaries in the catalog.

In Table \ref{tab:sample}, we confirm the results of \cite{berski16} and \cite{bobylev17} that HIP 89825 (GJ 710) is still the closest recent encounter of the Solar System. Unlike other close encounters of the Solar System, the perihelion of HIP 89825 is well determined due to the high precision TGAS astrometry. The perihelion of HIP 103738 is 0.93\,pc, consistent with the value given by \cite{dybczynski15} and by \cite{bailer-jones15}. 

Compared with the results in \cite{bailer-jones17} and \cite{bobylev17}, we only report encounters with perihelion less than 1\,pc and only use data with reliable radial velocity and astrometry. By including TAGS and RAVE targets with large measurement uncertainty in their sample, \cite{bailer-jones17} and \cite{bobylev17} find some TGAS targets with perihelion less than 1\,pc. But they admit that these encounters have unreliable data. Therefore, the TGAS astrometry do not lead to identification of new close encounters although it improves the precision of the perihelion of GJ 710. 

\begin{table}
\caption{Encounters of the Solar System and their nominal, mean, 5\% and 95\% quantiles of $d_{\rm enc}$, $t_{\rm enc}$, and $v_{\rm enc}$. The encounters are sorted by $d_{\rm enc}^{\rm nom}$. The names beginning with ``J'' are LAMOST IDs. The column ``cat.'' shows the catalogue we described in section \ref{sec:data}. }
\label{tab:sample}
\centering 
\begin{tabular}{lccccccccccccc}
  \hline
  Name &cat.&$d_{\rm enc}^{\rm nom}$&$d_{\rm enc}^{\rm mean}$ & $d_{\rm enc}^{\rm 5\%}$& $d_{\rm enc}^{\rm 95\%}$ &$t_{\rm enc}^{\rm nom}$&$t_{\rm enc}^{\rm mean}$ &$t_{\rm enc}^{\rm 5\%}$ &$t_{\rm enc}^{\rm 95\%}$ &$v_{\rm enc}^{\rm nom}$&$v_{\rm enc}^{\rm mean}$ &$v_{\rm enc}^{\rm 5\%}$ &$v_{\rm enc}^{\rm 95\%}$ \\
&  &(pc)&(pc)   &(pc)&(pc)&(Myr)&(Myr)&(Myr)&(Myr)&(km/s)&(km/s)&(km/s)&(km/s)\\
\hline
   HIP 89825 & xt & 0.063 & 0.063 & 0.038 & 0.089 & 1.35 & 1.35 & 1.31 & 1.41 & 13.80 & 13.80 & 13.30 & 14.26 \\ 
  HIP 71681 & rh & 0.898 & 0.900 & 0.819 & 0.987 & 0.03 & 0.03 & 0.03 & 0.03 & 30.77 & 30.82 & 28.99 & 32.55 \\ 
    HIP 103738 & xh & 0.929 & 0.919 & 0.442 & 1.528 & -3.85 & -3.86 & -4.18 & -3.57 & 17.97 & 17.94 & 16.67 & 19.20 \\ 
   HIP 71683 & xh & 0.976 & 0.976 & 0.947 & 1.007 & 0.03 & 0.03 & 0.03 & 0.03 & 31.63 & 31.62 & 30.71 & 32.54 \\ 
   \hline
\end{tabular}
\end{table}

\subsection{Extremely close encounter pairs}\label{sec:pair}
We also show encounter pairs with $d_{\rm enc}<0.02$\,pc in Table \ref{tab:ep0.02}. We see the periastrons of most extremely close encounter pairs have large errors with respect to their extremely small periastrons. Some of these extremely close encounter pairs are probably binaries since their initial positions are very close and the corresponding $t_{\rm enc}$ is close to 0. For example, HIP 41181 and HIP 41184 probably form a wide binary because they are currently close and the encounter time is 0.012\,Myr. We further describe these encounter pairs in detail based on $d_{\rm enc}^{\rm nom}$ as follows.
\begin{itemize}
\item  {\bf HIP 9974 and HIP 107038}. HIP 9974 (HD 13014) is an F5V star with a mass of 1.197$\pm$0.088\,$M_\odot$ \citep{grenier99} and metal rich ([Fe/H]=0.25, \citealt{robinson07}). While \cite{gontcharov06} derived a RV of -5.6$\pm$0.6 km/s, it has precision RV measurements in the Sophie archive \citep{perruchot11} indicating a long term RV variation of around 260~m/s per year caused by companion(s), consistent with \cite{makarov05} who find it having a 3.5 sigma discrepant proper motions between Hipparcos and Tycho. HIP 107038 (HD 207897) is a a moderately metal-poor ([Fe/H]=-0.14, \citealt{kotoneva02}) K3 star \citep{bidelman85} with a mass of 0.76\,$M_\odot$ \citep{brewer16}. 
\item  {\bf HIP 28898 and HIP 93373}. HIP 28898 (HD 42286) is a K1V star \citep{gray06} with a mass of 0.710$\pm$0.026 M\,$M_\odot$ (\citealt{mints17}, hereafter denoted by M17) it is listed as a possible subdwarf by \cite{jao09}. HIP 93373 (HD 175607) is the most metal-poor G dwarf ([Fe/H]=-0.65) with an orbiting sub-Neptune hosting a planet \citep{mortier16} and with a mass of 0.773$\pm$0.032\,$M_\odot$ (M17).
\item {\bf HIP 41181 and HIP 41184}. HIP 41181 (BD+45 1575a) and HIP 41184 (HD 70516) are a common proper motion pair \citep{gould04}. HIP 41181 is a metal poor ([Fe/H]=-0.54$\pm$0.07, \citealt{netopil17}) G0 star \citep{white07} with a mass of 0.943$\pm$0.078\,$M_\odot$. In contrast \cite{mello14} consider HIP41184 as a potential Solar analog. It lies just outside their two sigma colour similarity and photometric $T_{\rm eff}$ selection but emphasize that its close effective temperature (5753\,K) and metallicity ([Fe/H]=0.03$\pm$0.05, \citealt{netopil17}) mean that it should be further investigated to further quantify its strong likeness to the Sun. The very different metallicities of the pair determined by the same study (and elsewhere) appears to be robust support for an encounter interpretation rather than the coevality usually associated with common proper motion pairs.
\item {\bf HIP 51074 and HIP 85598}. HIP 51074 (HD 90926) is a mildly metal rich ([Fe/H]=0.13, \citealt{mena14}) G6V star \citep{houk75} with a mass of 0.928$\pm$0.040\,$M_\odot$ (M17) while HIP 85598 (HD 157802) is a metal poor ([Fe/H]=-0.10, \citealt{casagrande11}) F0-dD star \citep{renson09} with a 1.595$\pm$0.109\,$M_\odot$ (M17). It is noted as a potential binary having 3.5 sigma discrepant proper motions between Hipparcos and Tycho \citep{makarov05}.
\item {\bf HIP 52727 and HIP 58380}.  HIP 52727 (HD 93497) is a G5III and G2V binary with an estimated 138 year orbit \citep{malkov12}. HIP 58380 (HD 103975) is a slightly metal poor ([Fe/H]=-0.12$\pm$0.04 \citep{netopil17} F8.5VFe-0.5 star \citep{gray06} with a mass of 1.134$\pm$0.065\,$M_\odot$ (M17).
\item {\bf TYC 8793-520-1 and TYC 8793-962-1}. TYC 8793-520-1 is a K star \citep{pickles10} with [Fe/H]=0.07 \citep{kunder16} mass of 0.752$\pm$0.023\,$M_\odot$ (M17) while TYC 8793-962-1 (HD 202326) is a G3/5IV star \citep{houk75} and with a mass of 0.979$\pm$0.072\,$M_\odot$ (M17).
\item {\bf HIP 107618 and HIP 110014}. HIP 107618 (HD 207237) is a metal poor ([Fe/H]=-0.29$\pm$0.03, \citealt{netopil17}) G3V star \citep{houk88} with a mass of 1.136$\pm$0.067 (M17). It has four precision RVs from SOPHIE, suggesting a value of 52.276$\pm$0.003\,km/s which is consistent with the XHIP RV of 52.2$\pm$0.3\,km/s used in this work. HIP 110014 (HD 211380) is a less metal poor ([Fe/H]=-0.16$\pm$0.03, \citealt{netopil17}) F5V star \citep{houk88} with a mass of 1.169$\pm$0.055\,$M_\odot$.
\end{itemize}

\begin{table}
\centering
\caption{The 7 encounter pairs with $d_{\rm enc}<0.02$\,pc ordered by the right ascension of the primary of a pair. The column ``comp.'' are components of encounter pairs while 1 and 2 represent the primary and secondary components, respectively. The data for the primary component is not given. }
\label{tab:ep0.02}
\begin{tabular}{lcccccccccccccc}
  \hline
  Name &cat.&comp.&$d_{\rm enc}^{\rm nom}$&$d_{\rm enc}^{\rm mean}$ & $d_{\rm enc}^{\rm 5\%}$& $d_{\rm enc}^{\rm 95\%}$ &$t_{\rm enc}^{\rm nom}$&$t_{\rm enc}^{\rm mean}$ &$t_{\rm enc}^{\rm 5\%}$ &$t_{\rm enc}^{\rm 95\%}$ &$v_{\rm enc}^{\rm nom}$&$v_{\rm enc}^{\rm mean}$ &$v_{\rm enc}^{\rm 5\%}$ &$v_{\rm enc}^{\rm 95\%}$ \\
&&&(pc)&(pc)&(pc)&(pc)&(Myr)&(Myr)&(Myr)&(Myr)&(km/s)&(km/s)&(km/s)&(km/s)\\
  \hline
HIP 9974 & xh & 1 &  &  &  &  &  &  &  &  &  &  &  &  \\ 
  HIP 107038 & xt & 2 & 0.0168 & 0.09 & 0.00 & 2.5 & 1.196 & 1.198 & 1.087 & 1.329 & 34.6 & 34.5 & 33.4 & 35.7 \\ 
  HIP 28898 & xt & 1 &  &  &  &  &  &  &  &  &  &  &  &  \\ 
  HIP 93373 & xh & 2 & 0.0176 & 0.10 & 0.00 & 3.3 & 0.165 & 0.165 & 0.155 & 0.175 & 217.1 & 217.3 & 213.2 & 221.3 \\ 
  HIP 41181 & xt & 1 &  &  &  &  &  &  &  &  &  &  &  &  \\ 
  HIP 41184 & xt & 2 & 0.0043 & 0.06 & 0.00 & 0.8 & 0.012 & 0.079 & -0.251 & 0.666 & 0.6 & 0.8 & 0.4 & 1.3 \\ 
  HIP 51074 & xt & 1 &  &  &  &  &  &  &  &  &  &  &  &  \\ 
  HIP 85598 & xh & 2 & 0.0072 & 0.03 & 0.00 & 4.0 & -1.001 & -1.002 & -1.113 & -0.896 & 47.2 & 47.3 & 45.5 & 48.9 \\ 
  HIP 52727 & xh & 1 &  &  &  &  &  &  &  &  &  &  &  &  \\ 
  HIP 58380 & xt & 2 & 0.0163 & 0.04 & 0.00 & 1.4 & 0.204 & 0.204 & 0.200 & 0.209 & 36.8 & 36.8 & 36.1 & 37.5 \\ 
  TYC 8793-520-1 & rt & 1 &  &  &  &  &  &  &  &  &  &  &  &  \\ 
  TYC 8793-962-1 & rt & 2 & 0.0182 & 0.37 & 0.00 & 3.1 & -0.141 & -0.201 & -2.297 & 1.833 & 4.4 & 4.5 & 1.0 & 8.5 \\ 
  HIP 107618 & xt & 1 &  &  &  &  &  &  &  &  &  &  &  &  \\ 
  HIP 110014 & xt & 2 & 0.0159 & 0.005 & 0.00 & 1.8 & -0.099 & -0.099 & -0.110 & -0.087 & 64.9 & 64.9 & 64.1 & 65.8 \\ 
  \hline
\end{tabular}
\end{table}

\subsection{Encounters of planetary systems}\label{sec:companion}
Encounters can also perturb the outer debris disc of planetary systems and increase the impact rate, affecting the habitability of planets in the habitable zone \citep{kopparapu14}. For example, GJ 710 will pass the Sun at 0.06\,pc about 1.35\,Myr from the present, leading to an increase of cometary flux by at least one order of magnitude \citep{berski16}. In the CEP1 catalog, we find 577 encounters of known planetary systems. For the purpose of follow up studies, we show a sample of encounters of well-known planetary systems in Table \ref{tab:host}. For example, Epsilon Eridani is a nearby young K star with a debris disk and two suspected planets \citep{campbell88,greaves98,quillen02}. This system passed HIP 77158 and HIP 39342 at 0.3\,pc and will pass HIP 106255 at 0.28\,pc. A potential Oort cloud around Epsilon Eridani can be strongly perturbed by these encounters and trigger comet shower in the inner system. 

\begin{longrotatetable}
  \begin{deluxetable*}{lcccccccccccccccccc}
\tablecaption{A sample of close encounters of planet hosts. The parameters of the host stars are from \url{http://exoplanet.eu}. The names of encounters and corresponding encounter parameters are given by the ``encounter'' and subsequent columns. The columns to the left of the ``encounter'' column are the name, mass, radii, age, number of planets, observation of disc, and simbad spectral type of the planet hosts.}
\tablewidth{700pt}
\label{tab:host}
\tabletypesize{\scriptsize}
\tablehead{
 \colhead{Planet host}&\colhead{$M_*$}&\colhead{$R_*$}&\colhead{age}&\colhead{$N_p$}&\colhead{disc}&\colhead{Spectral type}&\colhead{encounter}&\colhead{cat.}&\colhead{$d_{\rm enc}^{\rm nom}$}& \colhead{$d_{\rm enc}^{\rm 5\%}$}&\colhead{ $d_{\rm enc}^{\rm 95\%}$} &\colhead{$t_{\rm enc}^{\rm nom}$}&\colhead{$t_{\rm enc}^{\rm 5\%}$} &\colhead{$t_{\rm enc}^{\rm 95\%}$} &\colhead{$v_{\rm enc}^{\rm nom}$} &\colhead{$v_{\rm enc}^{\rm 5\%}$} &\colhead{$v_{\rm enc}^{\rm 95\%}$}\\
 &\colhead{ ($M_\odot$)}&\colhead{($R_\odot$)}
 &\colhead{ (Gyr)}&\colhead{}&\colhead{}&\colhead{}&\colhead{}&\colhead{}&\colhead{ (pc)}&\colhead{ (pc)}&\colhead{ (pc)}&\colhead{ (Myr)}&\colhead{ (Myr)}&\colhead{ (Myr)}&\colhead{ (km/s)}&\colhead{ (km/s)}&\colhead{ (km/s)}
 }
 \startdata  
 eps Eridani & 0.83 & 0.90 & 0.66 &   1 & Imaging & K2V & HIP 77158 & xt & 0.30 & 0.22 & 0.40 & -0.56 & -0.57 & -0.55 & 69.22 & 68.71 & 69.74 \\ 
  eps Eridani & 0.83 & 0.90 & 0.66 &   1 & Imaging & K2V & HIP 39342 & xt & 0.30 & 0.23 & 0.38 & -0.34 & -0.35 & -0.33 & 48.45 & 47.14 & 49.81 \\ 
  eps Eridani & 0.83 & 0.90 & 0.66 &   1 & Imaging & K2V & HIP 106255 & xh & 0.28 & 0.24 & 0.91 & 0.11 & 0.10 & 0.12 & 78.46 & 76.86 & 80.33 \\ 
  GJ 785 & 0.78 & 0.68 &  &   2 & IR Excess & K1V & HIP 5643 & xh & 0.36 & 0.26 & 0.55 & 0.17 & 0.17 & 0.18 & 46.80 & 44.38 & 49.07 \\ 
HIP 58451& 0.79 &  & 4.33 &   1 & IR Excess & K2V & HIP 32673 & xh & 0.09 & 0.00 & 4.54 & -0.61 & -0.67 & -0.56 & 78.33 & 77.18 & 79.76 \\ 
HIP 58451& 0.79 &  & 4.33 &   1 & IR Excess & K2V & HIP 90539 & xt & 0.36 & 0.33 & 1.35 & 0.53 & 0.53 & 0.54 & 82.85 & 82.27 & 83.46 \\ 
  HIP 7978 & 1.10 & 1.10 & 4.80 &   1 & Imaging & F8V & HIP 110084 & xt & 0.45 & 0.42 & 1.29 & 0.38 & 0.37 & 0.40 & 93.25 & 92.52 & 94.01 \\ 
  HIP 8770 & 1.19 & 1.38 & 5.40 &   2 &  & G0V & HIP 7886 & xt & 0.09 & 0.00 & 2.48 & 0.44 & 0.41 & 0.46 & 38.47 & 37.34 & 39.94 \\ 
  HIP 1499 & 1.02 & 1.09 & 6.30 &   2 & IR Excess & G0V & HIP 42291 & xt & 0.37 & 0.12 & 1.41 & 0.76 & 0.75 & 0.78 & 53.53 & 52.35 & 54.77 \\ 
  HIP 85017 &  &  &  &   1 &  & G8.5V & HIP 83990 & xt & 0.08 & 0.07 & 0.38 & 0.19 & 0.19 & 0.20 & 123.41 & 121.86 & 124.82 \\ 
HIP 93373 & 0.71 & 0.71 & 10.32 &   1 &  & G6V & HIP 28898 & xt & 0.02 & 0.00 & 3.42 & 0.16 & 0.15 & 0.17 & 217.12 & 213.41 & 221.25 \\ 
  HIP 93017 & 0.90 &  & 8.1 &   1 & IR Excess & F9V & HIP 78519 & xt & 0.42 & 0.07 & 1.36 & -0.88 & -0.90 & -0.86 & 52.61 & 51.83 & 53.43 \\ 
HIP 97546 & 1.22 &  & 3.30 &   1 & IR Excess & G0V & HIP 1936 & xt & 0.18 & 0.12 & 0.73 & -0.59 & -0.61 & -0.57 & 65.36 & 64.37 & 66.36 \\ 
  HIP 15510 & 0.81 & 0.90 & 14.00 &   3 & IR Excess & G8V & HIP 103039 & xh & 0.35 & 0.20 & 0.50 & -0.05 & -0.06 & -0.05 & 140.71 & 140.12 & 141.35 \\ 
  HIP 109378 & 1.09 & 1.10 & 6.93 &   1 & IR Excess & G0V & HIP 1031 & xt & 0.41 & 0.08 & 0.76 & 0.33 & 0.33 & 0.33 & 58.49 & 58.13 & 58.91 \\ 
  HIP 20723& 1.24 & 1.03 & 5.36 &   1 &  & G6.5IV-V & TYC 6470-960-1 & rt & 0.07 & 0.00 & 1.98 & 0.93 & 0.87 & 0.98 & 53.91 & 52.65 & 55.31 \\ 
  HIP 33212 & 1.04 & 1.11 & 4.58 &   1 & IR Excess & F8V & HIP 40167 & xh & 0.31 & 0.15 & 1.29 & -0.99 & -1.06 & -0.92 & 10.64 & 10.26 & 11.05 \\ 
  HIP 33212& 1.04 & 1.11 & 4.58 &   1 & IR Excess & F8V & HIP 57050 & xt & 0.28 & 0.15 & 0.44 & 0.99 & 0.97 & 1.01 & 26.67 & 26.41 & 26.93 \\ 
  HIP 40693& 0.86 & 0.90 & 7.00 &   3 & IR Excess & K0V & HIP 70648 & xt & 0.44 & 0.16 & 2.37 & 0.91 & 0.89 & 0.94 & 75.08 & 74.43 & 75.75 \\ 
  HIP 54906 & 0.85 & 0.73 &  &   1 &  & K1V & HIP 50341 & xh & 0.06 & 0.06 & 0.83 & -0.30 & -0.32 & -0.28 & 46.00 & 43.37 & 48.85 \\ 
  Kapteyn's & 0.28 & 0.29 & 8.00 &   2 &  & M1VIp & HIP 99965 & xh & 0.99 & 0.93 & 1.10 & -0.13 & -0.13 & -0.13 & 257.31 & 256.84 & 257.77 \\ 
  Kapteyn's & 0.28 & 0.29 & 8.00 &   2 &  & M1VIp & HIP 110893 & xh & 0.87 & 0.85 & 0.89 & -0.03 & -0.03 & -0.03 & 265.23 & 264.71 & 265.73 \\ 
\enddata
\end{deluxetable*}
\end{longrotatetable}

\section{Potential applications of encounter pair catalogs}\label{sec:application}
From Table \ref{tab:ep0.02}, we see some encounter pairs which are less than 0.02\,pc apart from each other. These encounter pairs provide a real sample to study the formation and evolution of wide binaries. For example, a star can capture planets or disc objects from an encounter in its birth cluster \citep{levison10}. Wide companions can also be strongly perturbed and even ejected from the binaries by encounters and the Galactic tide \citep{feng17f, feng18}. Although simulated encounters are typically used to study the cumulative influence of encounters on stellar/planetary systems, a real sample of encounters can be used (1) to derive the encounter rate and thus to build a realistic encounter model, (2) to study the influence of individual encounters on the Oort-cloud-like structure of a system \citep{feng15b}, and (3) to study the influence of individual massive encounters on the dynamical stability of a system since massive close encounters are able to influence a system over a long time scale (e.g., \citealt{malmberg10,fouchard11}). 

The CEP1 catalog can be extended by including the orbits of star formation region star clusters. Stellar kinematics has been frequently used to trace the birth places of neutron stars \citep{tetzlaff09}, runaway stars \citep{boubert17,marchetti17}, and find solar siblings \citep{wielen96,martinez-barbosa16}. By identifying encounters between field stars with stars in a star formation region, one can find the birth place of the field star. Specifically, the age and astrometric information of stellar encounters and star formation region are needed to reconstruct their orbits and to diagnose whether they have similar age and are associated. If a stellar system encounters a massive star cluster or a molecular cloud, it would probably be captured by the cluster or disintegrated if it contains multiple components. Although it may appear that some stars encounter a cluster with relatively high speed, there will be many slower encounters which could be bound to the cluster in the past. For example, the mean encounter velocity $v_{\rm enc}$ of encounter pairs is about 65\,km/s according to Fig. \ref{fig:hist_dtv}. There are 10 encounters with periastron less than 1\,pc in the CEP1 catalog with $v_{\rm enc}<1$\,km/s. Considering the incompleteness of this catalog, the considerable errors in the the data and that the cross section of a star cluster is significantly larger than that of a single star, there could be thousands of stars which have escaped from a cluster in the FS catalog. With more precise astrometry for these stars, it would be meaningful to trace their orbits back to the escape epoch and to find whether they belonged to the cluster based on simulations. 

Moreover, the CEP1 catalog could be extended by including the orbits of supernova progenitors. The extended catalog could be used to estimate the position of supernova explosions and to find nearby planetary systems which were/will be influenced by these high-energy events (e.g. \citealt{breitschwerdt16}). For example, if a planetary system is less than 100\,pc away from a supernova, its habitability would be reduced either by direct damage of organisms or by secondary effects such as tropospheric ionization \citep{thomas16}. Recent studies of the $^{60}$Fe signature in the deep-sea crust on the Earth shows that the Earth has been less than 100\,pc from multiple supernova explosions in the past 2\,Myr \citep{breitschwerdt16,wallner16}. Hence we expect discoveries of closer encounters with supernova in the extended CEP1 catalog. In summary, CEP1 provides a real sample of encounters of binaries and planetary systems to study the influence of encounters on the dynamics and habitability of planets and companions. With a more comprehensive encounter catalogue based on future data releases, the forecast of ``interstellar climate'' for exoplanets would be feasible in the near future.

\begin{figure}
  \centering
      \includegraphics[scale=0.5]{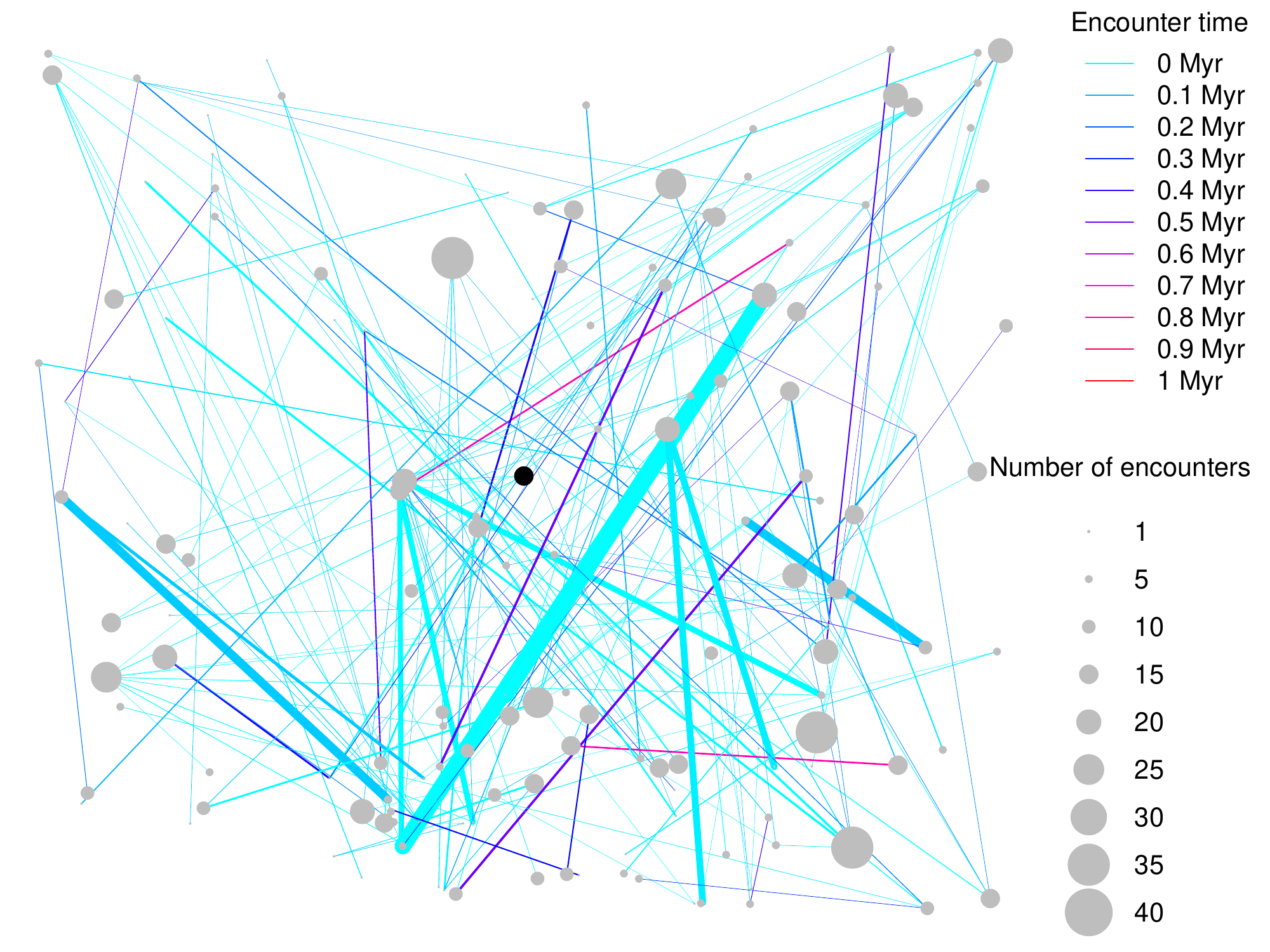}
      \caption{Network of stellar encounters in the solar neighborhood based on the EP1 catalog. The grey points represent stars less than 10 pc from the Sun in Galactic coordinates. The small black dot at the center denotes the Sun. The size of a dot is proportional to the number of encounters of a star. The lines connect encounter pairs which will pass each other at less than 1 pc. The width of the line is inversely proportional to the periastron of an encounter pair and the color of a line represents the encounter time.}
  \label{fig:net}
\end{figure}

As mentioned in Section \ref{sec:introduction}, the sample of encounter pairs also provides a time-varying network of stars, planets and thus potentially habitable worlds, which is illustrated by the stars within 10\,pc of the Sun in Fig. \ref{fig:net}. Considering that this network suffers from incompletenesses, we generate random networks to derive the indicative time-scale for interstellar travel, and design the following two exploration strategies:
\begin{itemize}
  \item Strategy 1 (S1, or trivial strategy): Humans launch one manned-spaceship every 100 years to reach the nearest habitable world and travel with a constant speed. If the nearest world cannot be reached within 100 years, humans wait for another 100 year to find possible explorable worlds. The explored worlds also adopt this strategy to explore other worlds. 
  \item Strategy 2 (S2, or optimization strategy): Similar to the first strategy, but for exploration in an optimized fashion. Humans explore the other worlds by optimizing launch time and choosing target worlds which are near-by and with relative velocity such that the exploration time can be reduced. 
  \end{itemize}

To derive the exploration time scale, we (1) draw a sample of nearby stars with velocities simulated using the model in \cite{feng14} for a given star according to a number density of 0.25\,pc$^{-3}$ \citep{feng18}, and randomly select habitable worlds from them by adopting a ratio of habitability worlds $f_h$ (or habitability ratio), (2) select the next explorable candidate based on the first or second strategy, (3) calculate the travel time, (4) start a new exploration by repeating the first three steps every 100\,year for 10 times, and (5) repeat step (1) to (4) for 10 times to estimate the uncertainty. Considering the current technology and humans' life span, we fix the launch rate at one spaceship per 100\,year and the maximum travel time at 100\,year. We vary the travel speed $v_t$ from 0.01$c$ to 1$c$, where $c$ is the speed of light. Although the occurrence rate of habitable zone planets have been intensively investigated \citep{dressing15,burke15}, the ratio of habitable worlds are poorly known due to a lack of reliable criterion for habitability assessment. 

We also vary the habitability ratio $f_h$ from 0.1 to 1. For both strategies, we find that the travel speed is crucial and interstellar travel is only possible if the speed is larger than about 0.01$c$. This is due to the fact that the number of explorable habitable worlds for a given limit of travel time $T$ is proportional to $f_h(v_tT)^{3}$. Thus the explorable worlds are limited mainly by the travel speed. For travel speed higher than 0.4$c$, the difference in the travel efficiency between different strategies disappears because the spaceship is much faster than stellar motions which are used to reduce travel time.

\begin{figure}
  \centering
      \includegraphics[scale=0.7]{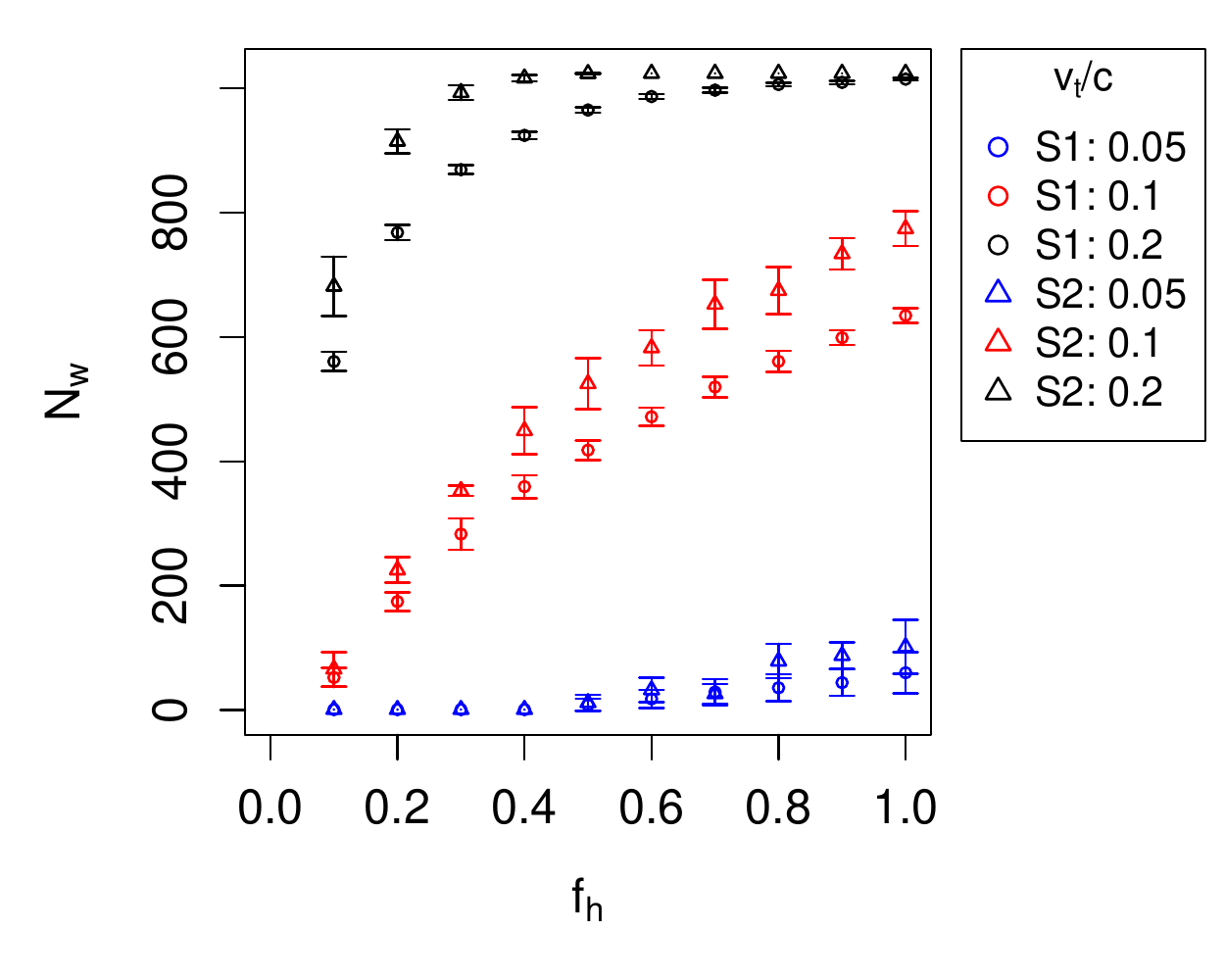}
      \includegraphics[scale=0.7]{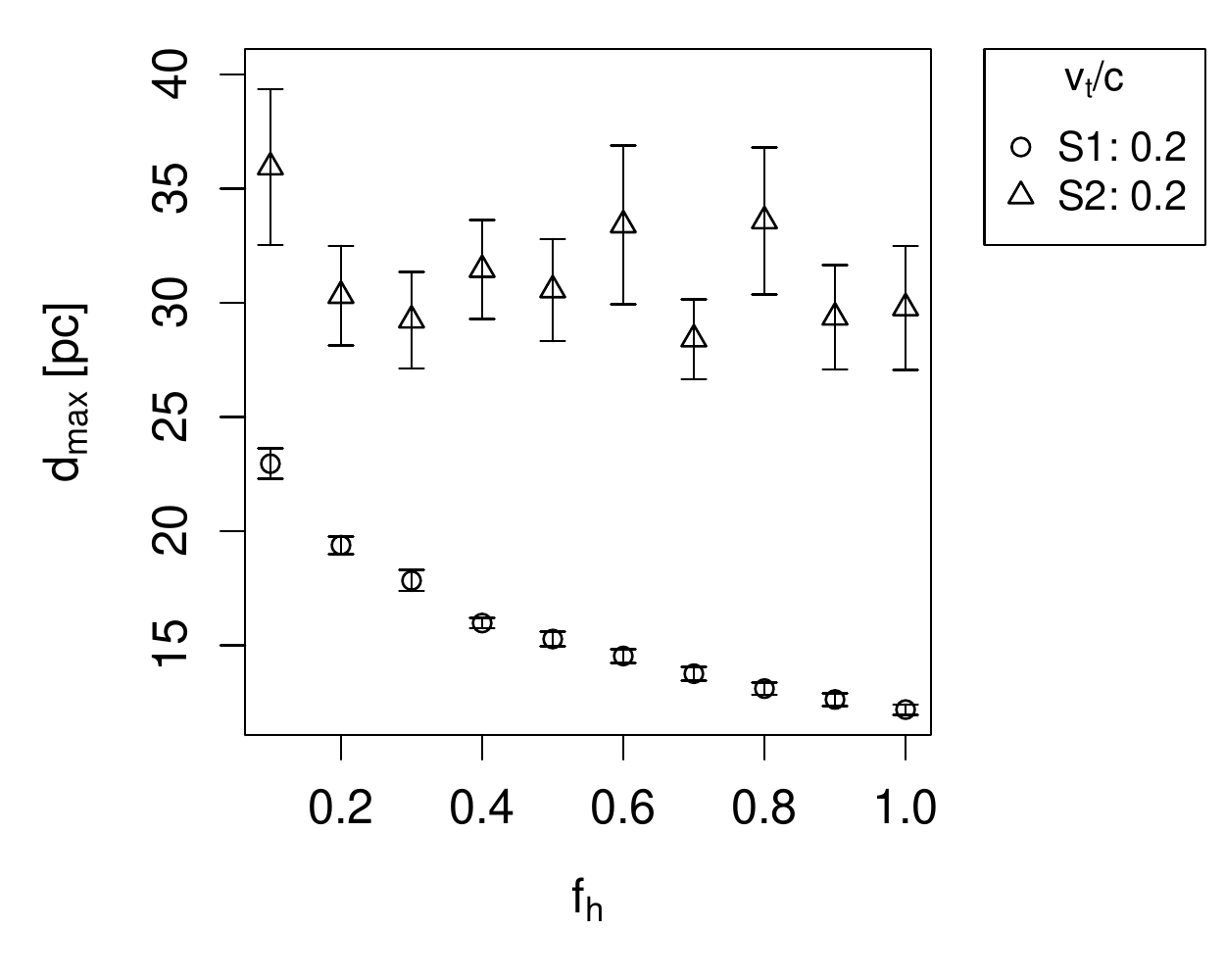}
  \caption{Comparison of the exploration efficiency for S1 and S2. Left panel: the number of explored worlds $N_w$ as a function of $f_h$ and $v_t$. Right panel: the distance of the farthest explored worlds as a function of $f_h$ for $v_t=0.2$c. }
  \label{fig:travel}
\end{figure}

We show the number of explored worlds $N_w$ for S1 and S2 as a function of $v_t$ and $f_h$ in Figure \ref{fig:travel}. We find that an optimization strategy is more important for exploration of habitable worlds for $v_t=0.1c$ than for $v_t=0.05c$ and $v_t=0.2c$. Optimization is important for high habitability ratio if the travel speed is low while it is important for low habitability ratio if the travel speed is high. Moreover, S2 can be used to explore more distant worlds than S1, as seen in the right panel of Figure \ref{fig:travel}. The maximum distance explored based on S1 decrease with $f_h$ because the habitable worlds are scarce and far away from a given world for a low habitability ratio.

For most cases that we investigate S2 reduces the time to explore distant worlds and becomes particularly important when the habitability ratio and travel speed are low and less habitable worlds are available to use as a stepping stone for onward travel. We caution that our brief investigation of interstellar travel is relatively simplistic and does not account for many important factors, e.g., time-varying travel speed and the different habitability levels of planetary systems. We expect that a more comprehensive study on interstellar travel will become appropriate with forthcoming improvements in the quality of astrometric and radial velocity data supplemented by increasingly robust metrics for quantifying  planetary habitability.

\section{Conclusion}\label{sec:conclusion}
We provide a catalogue of encounter pairs, including the astrometry and periastron parameters of 8149 encounter pairs based on simulations of the motions of 349,161 stars with initial conditions derived mainly from the TGAS, Hipparcos and RAVE5 catalogues. We find 577 encounters of stars which host planetary systems and/or debris disks. We find 4 encounters of the Solar System with perihelion less than 1\,pc and 96 encounters with periastron less than 0.1\,pc. Since we only use the most reliable data, there is significant incompleteness in the sample of encounters (e.g., for the Solar encounters see the middle panel of Fig. \ref{fig:hist_dtv}). 

We find significant anisotropy in the angular distribution of encounter pairs. This anisotropy is caused by the solar apex motion, which is added onto the helio-static motions of stars for orbital integrations. This observed anisotropy is consistent with the prediction of \cite{feng14} based on simulated encounters. The anisotropic encounters would perturb planetary systems preferentially and induce a tidal plane, where the tidal force is stronger. Similar to the Kozai mechanism, this mechanism would probably change the orbits of wide companions and outer planets in a stellar system, which can further influence the habitability of inner planets through the Kozai mechanism \citep{kaib13}.

The CEP1 catalogue can be used as a starting point to study the dynamics of nearby hierarchical stellar systems such as Fomalhaut, alpha Centauri, and to identify non-coeval wide binary candidates such as HIP 41181 and HIP 41184. This catalog can be extended by integrating the orbits of molecular clouds, supernova progenitor, and star clusters. The extended catalog can be applied to identify the birth place of stars and to study the influence of globular clusters, supernovae, and molecular clouds on the dynamics of stellar systems and on the habitability of planetary systems. Since a number of the stars in the CEP1 catalog are known to host planets, it also provides encounters of planetary systems, showing the potential to forecast and/or reconstruct the interstellar climate of planetary systems.

Since this encounter catalog provides time-varying stellar neighbors, potentially habitable planets around these neighbors would form a network of habitable worlds. By simulating networks of habitable worlds, we find that interstellar travel can be optimized based on the use of motions of the worlds in the network. Given a low occurrence rate of habitable worlds and travel speed, an optimized strategy is crucial for the success of humans' interstellar travel. This motivates the importance of identifying a more complete network of stellar encounters as well as the detection and better quantification of potentially habitable planets around nearby stars. 

Most encounter pairs in the CEP1 catalogue are not determined with precise astrometry. With the upcoming Gaia data releases, we expect a significant improvement in the precision of periapses for a much larger sample. In addition, the Galactic potential in our work is not determined in a self-consistent way because our sample is relatively local to the Sun and thus cannot be used to constrain the global model of the Galaxy.

\section*{Acknowledgement}
We thank Coryn Bailer-Jones for valuable discussions during the initial investigation of this project. We also thank Niall Deacon for pointing out important references. The comments from anonymous referee enable a very significant improvement of the manuscript. The research is funded by the Science and Technology Facilities Council (ST/M001008/1). We have used the exoplanet catalog provided by \url{http://exoplanet.eu/catalog/}, and used Simbad and VizieR at the CDS website to obtain and cross-match data sets. 
\bibliographystyle{aasjournal}
\bibliography{nm}

\begin{thebibliography}{}
\expandafter\ifx\csname natexlab\endcsname\relax\def\natexlab#1{#1}\fi
\providecommand{\url}[1]{\href{#1}{#1}}

\bibitem[{{Altmann} {et~al.}(2017){Altmann}, {Roeser}, {Demleitner}, {Bastian},
  \& {Schilbach}}]{altmann17}
{Altmann}, M., {Roeser}, S., {Demleitner}, M., {Bastian}, U., \& {Schilbach},
  E. 2017, \aap, 600, L4

\bibitem[{{Anderson} \& {Francis}(2012)}]{anderson11}
{Anderson}, E., \& {Francis}, C. 2012, Astronomy Letters, 38, 331

\bibitem[{{Andrews} {et~al.}(2017){Andrews}, {Chanam{\'e}}, \&
  {Ag{\"u}eros}}]{andrews17}
{Andrews}, J.~J., {Chanam{\'e}}, J., \& {Ag{\"u}eros}, M.~A. 2017, ArXiv
  e-prints, arXiv:1704.07829

\bibitem[{{Bailer-Jones}(2009)}]{bailer-jones09}
{Bailer-Jones}, C.~A.~L. 2009, Int. J. Astrob., 8, 213

\bibitem[{{Bailer-Jones}(2015)}]{bailer-jones15}
---. 2015, \aap, 575, A35

\bibitem[{{Bailer-Jones}(2017)}]{bailer-jones17}
---. 2017, A{\&}A, accepted

\bibitem[{{Batygin} \& {Brown}(2016)}]{batygin16}
{Batygin}, K., \& {Brown}, M.~E. 2016, \aj, 151, 22

\bibitem[{{Berski} \& {Dybczy{\'n}ski}(2016)}]{berski16}
{Berski}, F., \& {Dybczy{\'n}ski}, P.~A. 2016, \aap, 595, L10

\bibitem[{{Bidelman}(1985)}]{bidelman85}
{Bidelman}, W.~P. 1985, \apjs, 59, 197

\bibitem[{{Binney} \& {Tremaine}(2008)}]{binney08_book}
{Binney}, J., \& {Tremaine}, S. 2008, {Galactic Dynamics: Second Edition}
  (Princeton University Press)

\bibitem[{{Bobylev} \& {Bajkova}(2017)}]{bobylev17}
{Bobylev}, V.~V., \& {Bajkova}, A.~T. 2017, Astronomy Letters, 43, 559

\bibitem[{{Boubert} {et~al.}(2017){Boubert}, {Erkal}, {Evans}, \&
  {Izzard}}]{boubert17}
{Boubert}, D., {Erkal}, D., {Evans}, N.~W., \& {Izzard}, R.~G. 2017, \mnras,
  469, 2151

\bibitem[{{Breitschwerdt} {et~al.}(2016){Breitschwerdt}, {Feige}, {Schulreich},
  {Avillez}, {Dettbarn}, \& {Fuchs}}]{breitschwerdt16}
{Breitschwerdt}, D., {Feige}, J., {Schulreich}, M.~M., {et~al.} 2016, \nat,
  532, 73

\bibitem[{{Brewer} {et~al.}(2016){Brewer}, {Fischer}, {Valenti}, \&
  {Piskunov}}]{brewer16}
{Brewer}, J.~M., {Fischer}, D.~A., {Valenti}, J.~A., \& {Piskunov}, N. 2016,
  \apjs, 225, 32

\bibitem[{{Burke} {et~al.}(2015){Burke}, {Christiansen}, {Mullally}, {Seader},
  {Huber}, {Rowe}, {Coughlin}, {Thompson}, {Catanzarite}, {Clarke}, {Morton},
  {Caldwell}, {Bryson}, {Haas}, {Batalha}, {Jenkins}, {Tenenbaum}, {Twicken},
  {Li}, {Quintana}, {Barclay}, {Henze}, {Borucki}, {Howell}, \&
  {Still}}]{burke15}
{Burke}, C.~J., {Christiansen}, J.~L., {Mullally}, F., {et~al.} 2015, \apj,
  809, 8

\bibitem[{{Campbell} {et~al.}(1988){Campbell}, {Walker}, \&
  {Yang}}]{campbell88}
{Campbell}, B., {Walker}, G.~A.~H., \& {Yang}, S. 1988, \apj, 331, 902

\bibitem[{{Casagrande} {et~al.}(2011){Casagrande}, {Sch{\"o}nrich}, {Asplund},
  {Cassisi}, {Ram{\'{\i}}rez}, {Mel{\'e}ndez}, {Bensby}, \&
  {Feltzing}}]{casagrande11}
{Casagrande}, L., {Sch{\"o}nrich}, R., {Asplund}, M., {et~al.} 2011, \aap, 530,
  A138

\bibitem[{{Chandrasekhar}(1943)}]{chandrasekhar43}
{Chandrasekhar}, S. 1943, \apj, 97, 255

\bibitem[{{Crifo} {et~al.}(2017){Crifo}, {Soubiran}, {Jasniewicz}, {Katz},
  {Sartoretti}, \& {Panuzzo}}]{crifo17}
{Crifo}, F., {Soubiran}, C., {Jasniewicz}, G., {et~al.} 2017, \aap, 601, L6

\bibitem[{{Dehnen} \& {Binney}(1998)}]{dehnen98b}
{Dehnen}, W., \& {Binney}, J.~J. 1998, \mnras, 298, 387

\bibitem[{{Delgado Mena} {et~al.}(2015){Delgado Mena}, {Bertr{\'a}n de Lis},
  {Adibekyan}, {Sousa}, {Figueira}, {Mortier}, {Gonz{\'a}lez Hern{\'a}ndez},
  {Tsantaki}, {Israelian}, \& {Santos}}]{mena14}
{Delgado Mena}, E., {Bertr{\'a}n de Lis}, S., {Adibekyan}, V.~Z., {et~al.}
  2015, \aap, 576, A69

\bibitem[{{Dhital} {et~al.}(2012){Dhital}, {West}, {Stassun}, {Bochanski},
  {Massey}, \& {Bastien}}]{dhital11}
{Dhital}, S., {West}, A.~A., {Stassun}, K.~G., {et~al.} 2012, \aj, 143, 67

\bibitem[{{Domainko} {et~al.}(2013){Domainko}, {Bailer-Jones}, \&
  {Feng}}]{domainko13}
{Domainko}, W., {Bailer-Jones}, C.~A.~L., \& {Feng}, F. 2013, \mnras, 432, 258

\bibitem[{{Dressing} \& {Charbonneau}(2015)}]{dressing15}
{Dressing}, C.~D., \& {Charbonneau}, D. 2015, \apj, 807, 45

\bibitem[{{Dybczy{\'n}ski} \& {Berski}(2015)}]{dybczynski15}
{Dybczy{\'n}ski}, P.~A., \& {Berski}, F. 2015, \mnras, 449, 2459

\bibitem[{{Feng} \& {Bailer-Jones}(2013)}]{feng13}
{Feng}, F., \& {Bailer-Jones}, C.~A.~L. 2013, \apj, 768, 152

\bibitem[{{Feng} \& {Bailer-Jones}(2014)}]{feng14}
---. 2014, \mnras, 442, 3653

\bibitem[{{Feng} \& {Bailer-Jones}(2015)}]{feng15b}
---. 2015, \mnras, 454, 3267

\bibitem[{{Feng} \& {Jones}(2017)}]{feng17f}
{Feng}, F., \& {Jones}, H.~R.~A. 2017, ArXiv e-prints, arXiv:1710.07184

\bibitem[{{Feng} \& {Jones}(2018)}]{feng18}
---. 2018, \mnras, 473, 3185

\bibitem[{{Finch} \& {Zacharias}(2016)}]{finch16}
{Finch}, C.~T., \& {Zacharias}, N. 2016, VizieR Online Data Catalog, 1333

\bibitem[{{Fouchard} {et~al.}(2011){Fouchard}, {Froeschl{\'e}}, {Rickman}, \&
  {Valsecchi}}]{fouchard11}
{Fouchard}, M., {Froeschl{\'e}}, C., {Rickman}, H., \& {Valsecchi}, G.~B. 2011,
  \icarus, 214, 334

\bibitem[{{Gaia Collaboration} {et~al.}(2016){Gaia Collaboration}, {Brown},
  {Vallenari}, {Prusti}, {de Bruijne}, {Mignard}, {Drimmel}, {Babusiaux},
  {Bailer-Jones}, {Bastian}, \& et~al.}]{gaia16}
{Gaia Collaboration}, {Brown}, A.~G.~A., {Vallenari}, A., {et~al.} 2016, \aap,
  595, A2

\bibitem[{{Garc{\'{\i}}a-S{\'a}nchez}
  {et~al.}(2001){Garc{\'{\i}}a-S{\'a}nchez}, {Weissman}, {Preston}, {Jones},
  {Lestrade}, {Latham}, {Stefanik}, \& {Paredes}}]{sanchez01}
{Garc{\'{\i}}a-S{\'a}nchez}, J., {Weissman}, P.~R., {Preston}, R.~A., {et~al.}
  2001, \aap, 379, 634

\bibitem[{Gizis {et~al.}(2001)Gizis, Kirkpatrick, Burgasser, Reid, Monet,
  Liebert, \& Wilson}]{gizis01}
Gizis, J.~E., Kirkpatrick, J.~D., Burgasser, A., {et~al.} 2001, The
  Astrophysical Journal Letters, 551, L163.
\newblock \url{http://stacks.iop.org/1538-4357/551/i=2/a=L163}

\bibitem[{{Gontcharov}(2006)}]{gontcharov06}
{Gontcharov}, G.~A. 2006, Astronomy Letters, 32, 759

\bibitem[{{Gould} \& {Chanam{\'e}}(2004)}]{gould04}
{Gould}, A., \& {Chanam{\'e}}, J. 2004, \apjs, 150, 455

\bibitem[{{Gray} {et~al.}(2006){Gray}, {Corbally}, {Garrison}, {McFadden},
  {Bubar}, {McGahee}, {O'Donoghue}, \& {Knox}}]{gray06}
{Gray}, R.~O., {Corbally}, C.~J., {Garrison}, R.~F., {et~al.} 2006, \aj, 132,
  161

\bibitem[{{Greaves} {et~al.}(1998){Greaves}, {Holland}, {Moriarty-Schieven},
  {Jenness}, {Dent}, {Zuckerman}, {McCarthy}, {Webb}, {Butner}, {Gear}, \&
  {Walker}}]{greaves98}
{Greaves}, J.~S., {Holland}, W.~S., {Moriarty-Schieven}, G., {et~al.} 1998,
  \apjl, 506, L133

\bibitem[{{Grenier} {et~al.}(1999){Grenier}, {Baylac}, {Rolland}, {Burnage},
  {Arenou}, {Briot}, {Delmas}, {Duflot}, {Genty}, {G{\'o}mez}, {Halbwachs},
  {Marouard}, {Oblak}, \& {Sellier}}]{grenier99}
{Grenier}, S., {Baylac}, M.-O., {Rolland}, L., {et~al.} 1999, \aaps, 137, 451

\bibitem[{{Heisler} {et~al.}(1987){Heisler}, {Tremaine}, \&
  {Alcock}}]{heisler87}
{Heisler}, J., {Tremaine}, S., \& {Alcock}, C. 1987, \icarus, 70, 269

\bibitem[{{Heller} \& {Hippke}(2017)}]{heller17}
{Heller}, R., \& {Hippke}, M. 2017, \apjl, 835, L32

\bibitem[{{H{\o}g} {et~al.}(2000){H{\o}g}, {Fabricius}, {Makarov}, {Urban},
  {Corbin}, {Wycoff}, {Bastian}, {Schwekendiek}, \& {Wicenec}}]{hog00}
{H{\o}g}, E., {Fabricius}, C., {Makarov}, V.~V., {et~al.} 2000, \aap, 355, L27

\bibitem[{{Houk} \& {Cowley}(1975)}]{houk75}
{Houk}, N., \& {Cowley}, A.~P. 1975, {University of Michigan Catalogue of
  two-dimensional spectral types for the HD stars. Volume I. Declinations
  -90{\deg} to -53{\deg}.}

\bibitem[{{Houk} \& {Smith-Moore}(1988)}]{houk88}
{Houk}, N., \& {Smith-Moore}, M. 1988, {Michigan Catalogue of Two-dimensional
  Spectral Types for the HD Stars. Volume 4, Declinations -26{\deg}.0 to
  -12{\deg}.0.}

\bibitem[{{Jao} {et~al.}(2009){Jao}, {Mason}, {Hartkopf}, {Henry}, \&
  {Ramos}}]{jao09}
{Jao}, W.-C., {Mason}, B.~D., {Hartkopf}, W.~I., {Henry}, T.~J., \& {Ramos},
  S.~N. 2009, \aj, 137, 3800

\bibitem[{{J{\'{\i}}lkov{\'a}} {et~al.}(2015){J{\'{\i}}lkov{\'a}}, {Portegies
  Zwart}, {Pijloo}, \& {Hammer}}]{jilkova15}
{J{\'{\i}}lkov{\'a}}, L., {Portegies Zwart}, S., {Pijloo}, T., \& {Hammer}, M.
  2015, \mnras, 453, 3157

\bibitem[{{Kaib} {et~al.}(2013){Kaib}, {Raymond}, \& {Duncan}}]{kaib13}
{Kaib}, N.~A., {Raymond}, S.~N., \& {Duncan}, M. 2013, \nat, 493, 381

\bibitem[{{Kopparapu} {et~al.}(2014){Kopparapu}, {Ramirez}, {SchottelKotte},
  {Kasting}, {Domagal-Goldman}, \& {Eymet}}]{kopparapu14}
{Kopparapu}, R.~K., {Ramirez}, R.~M., {SchottelKotte}, J., {et~al.} 2014,
  \apjl, 787, L29

\bibitem[{{Kotoneva} {et~al.}(2002){Kotoneva}, {Flynn}, {Chiappini}, \&
  {Matteucci}}]{kotoneva02}
{Kotoneva}, E., {Flynn}, C., {Chiappini}, C., \& {Matteucci}, F. 2002, \mnras,
  336, 879

\bibitem[{{Kozai}(1962)}]{kozai62}
{Kozai}, Y. 1962, \aj, 67, 591

\bibitem[{{Kunder} {et~al.}(2017){Kunder}, {Kordopatis}, {Steinmetz},
  {Zwitter}, {McMillan}, {Casagrande}, {Enke}, {Wojno}, {Valentini},
  {Chiappini}, {Matijevi{\v c}}, {Siviero}, {de Laverny}, {Recio-Blanco},
  {Bijaoui}, {Wyse}, {Binney}, {Grebel}, {Helmi}, {Jofre}, {Antoja}, {Gilmore},
  {Siebert}, {Famaey}, {Bienaym{\'e}}, {Gibson}, {Freeman}, {Navarro},
  {Munari}, {Seabroke}, {Anguiano}, {{\v Z}erjal}, {Minchev}, {Reid},
  {Bland-Hawthorn}, {Kos}, {Sharma}, {Watson}, {Parker}, {Scholz}, {Burton},
  {Cass}, {Hartley}, {Fiegert}, {Stupar}, {Ritter}, {Hawkins}, {Gerhard},
  {Chaplin}, {Davies}, {Elsworth}, {Lund}, {Miglio}, \& {Mosser}}]{kunder16}
{Kunder}, A., {Kordopatis}, G., {Steinmetz}, M., {et~al.} 2017, \aj, 153, 75

\bibitem[{{Lawson} {et~al.}(2016){Lawson}, {Murphy}, {Tinney}, {Ireland}, \&
  {Bessell}}]{lawson16}
{Lawson}, W., {Murphy}, S., {Tinney}, C.~G., {Ireland}, M., \& {Bessell}, M.~S.
  2016, in American Astronomical Society Meeting Abstracts, Vol. 228, American
  Astronomical Society Meeting Abstracts, 217.08

\bibitem[{{Levison} {et~al.}(2010){Levison}, {Duncan}, {Brasser}, \&
  {Kauffmann}}]{levison10}
{Levison}, H.~F., {Duncan}, M.~J., {Brasser}, R., \& {Kauffmann}, D.~E. 2010,
  in Bulletin of the American Astronomical Society, Vol.~42, AAS/Division of
  Dynamical Astronomy Meeting \#41, 925

\bibitem[{{Lidov}(1962)}]{lidov62}
{Lidov}, M.~L. 1962, \planss, 9, 719

\bibitem[{{Lingam} \& {Loeb}(2017)}]{lingam17}
{Lingam}, M., \& {Loeb}, A. 2017, ArXiv e-prints, arXiv:1703.00878

\bibitem[{{Luo} {et~al.}(2016){Luo}, {Zhao}, {Zhao}, {Deng}, {Liu}, {Jing},
  {Wang}, {Zhang}, {Shi}, {Cui}, {Chu}, {Li}, {Bai}, {Wu}, {Cai}, {Cao}, {Cao},
  {Carlin}, {Chen}, {Chen}, {Chen}, {Chen}, {Chen}, {Chen}, {Chen},
  {Christlieb}, {Chu}, {Cui}, {Dong}, {Du}, {Fan}, {Feng}, {Fu}, {Gao}, {Gong},
  {Gu}, {Guo}, {Han}, {He}, {Hou}, {Hou}, {Hou}, {Hu}, {Hu}, {Hu}, {Huo},
  {Jia}, {Jiang}, {Jiang}, {Jiang}, {Jin}, {Kong}, {Kong}, {Lei}, {Li}, {Li},
  {Li}, {Li}, {Li}, {Li}, {Li}, {Li}, {Li}, {Li}, {Li}, {Li}, {Liang}, {Lin},
  {Liu}, {Liu}, {Liu}, {Liu}, {Lu}, {Luo}, {Mao}, {Newberg}, {Ni}, {Qi}, {Qi},
  {Shen}, {Shi}, {Song}, {Song}, {Su}, {Su}, {Tang}, {Tao}, {Tian}, {Wang},
  {Wang}, {Wang}, {Wang}, {Wang}, {Wang}, {Wang}, {Wang}, {Wang}, {Wang},
  {Wang}, {Wang}, {Wang}, {Wang}, {Wang}, {Wang}, {Wang}, {Wang}, {Wang},
  {Wang}, {Wei}, {Wei}, {Wu}, {Wu}, {Wu}, {Wu}, {Xing}, {Xu}, {Xu}, {Xu},
  {Yan}, {Yang}, {Yang}, {Yang}, {Yang}, {Yao}, {Yu}, {Yuan}, {Yuan}, {Yuan},
  {Yuan}, {Zhai}, {Zhang}, {Zhang}, {Zhang}, {Zhang}, {Zhang}, {Zhang},
  {Zhang}, {Zhang}, {Zhao}, {Zhou}, {Zhou}, {Zhu}, {Zhu}, {Zou}, \&
  {Zuo}}]{luo16}
{Luo}, A.-L., {Zhao}, Y.-H., {Zhao}, G., {et~al.} 2016, VizieR Online Data
  Catalog, 5149

\bibitem[{{Majaess} {et~al.}(2009){Majaess}, {Turner}, \& {Lane}}]{majaess09}
{Majaess}, D.~J., {Turner}, D.~G., \& {Lane}, D.~J. 2009, \mnras, 398, 263

\bibitem[{{Makarov} \& {Kaplan}(2005)}]{makarov05}
{Makarov}, V.~V., \& {Kaplan}, G.~H. 2005, \aj, 129, 2420

\bibitem[{{Malkov} {et~al.}(2012){Malkov}, {Tamazian}, {Docobo}, \&
  {Chulkov}}]{malkov12}
{Malkov}, O.~Y., {Tamazian}, V.~S., {Docobo}, J.~A., \& {Chulkov}, D.~A. 2012,
  \aap, 546, A69

\bibitem[{{Malmberg} {et~al.}(2011){Malmberg}, {Davies}, \&
  {Heggie}}]{malmberg10}
{Malmberg}, D., {Davies}, M.~B., \& {Heggie}, D.~C. 2011, \mnras, 411, 859

\bibitem[{{Malmberg} {et~al.}(2007){Malmberg}, {de Angeli}, {Davies}, {Church},
  {Mackey}, \& {Wilkinson}}]{malmberg07}
{Malmberg}, D., {de Angeli}, F., {Davies}, M.~B., {et~al.} 2007, \mnras, 378,
  1207

\bibitem[{{Mamajek} {et~al.}(2015){Mamajek}, {Barenfeld}, {Ivanov}, {Kniazev},
  {V{\"a}is{\"a}nen}, {Beletsky}, \& {Boffin}}]{mamajek15}
{Mamajek}, E.~E., {Barenfeld}, S.~A., {Ivanov}, V.~D., {et~al.} 2015, \apjl,
  800, L17

\bibitem[{{Marchetti} {et~al.}(2017){Marchetti}, {Rossi}, {Kordopatis},
  {Brown}, {Rimoldi}, {Starkenburg}, {Youakim}, \& {Ashley}}]{marchetti17}
{Marchetti}, T., {Rossi}, E.~M., {Kordopatis}, G., {et~al.} 2017, ArXiv
  e-prints, arXiv:1704.07990

\bibitem[{{Mart{\'{\i}}nez-Barbosa}(2016)}]{martinez-barbosa16}
{Mart{\'{\i}}nez-Barbosa}, C.~A. 2016, PhD thesis, Leiden Observatory, Leiden
  University, P.B.~9513, Leiden NL-2300 RA, the Netherlands
  <EMAIL>cmartinez@strw.leidenuniv.nl</EMAIL>

\bibitem[{{Michalik} {et~al.}(2014){Michalik}, {Lindegren}, {Hobbs}, \&
  {Lammers}}]{michalik14}
{Michalik}, D., {Lindegren}, L., {Hobbs}, D., \& {Lammers}, U. 2014, \aap, 571,
  A85

\bibitem[{{Mints} \& {Hekker}(2017)}]{mints17}
{Mints}, A., \& {Hekker}, S. 2017, \aap, 604, A108

\bibitem[{{Monet} {et~al.}(2003){Monet}, {Levine}, {Canzian}, {Ables}, {Bird},
  {Dahn}, {Guetter}, {Harris}, {Henden}, {Leggett}, {Levison}, {Luginbuhl},
  {Martini}, {Monet}, {Munn}, {Pier}, {Rhodes}, {Riepe}, {Sell}, {Stone},
  {Vrba}, {Walker}, {Westerhout}, {Brucato}, {Reid}, {Schoening}, {Hartley},
  {Read}, \& {Tritton}}]{monet03}
{Monet}, D.~G., {Levine}, S.~E., {Canzian}, B., {et~al.} 2003, \aj, 125, 984

\bibitem[{{Mortier} {et~al.}(2016){Mortier}, {Faria}, {Santos}, {Rajpaul},
  {Figueira}, {Boisse}, {Collier Cameron}, {Dumusque}, {Lo Curto}, {Lovis},
  {Mayor}, {Melo}, {Pepe}, {Queloz}, {Santerne}, {S{\'e}gransan}, {Sousa},
  {Sozzetti}, \& {Udry}}]{mortier16}
{Mortier}, A., {Faria}, J.~P., {Santos}, N.~C., {et~al.} 2016, \aap, 585, A135

\bibitem[{{Mustill} {et~al.}(2016){Mustill}, {Raymond}, \&
  {Davies}}]{mustill16}
{Mustill}, A.~J., {Raymond}, S.~N., \& {Davies}, M.~B. 2016, \mnras, 460, L109

\bibitem[{{Napier}(2004)}]{napier04}
{Napier}, W.~M. 2004, \mnras, 348, 46

\bibitem[{{Netopil} {et~al.}(2017){Netopil}, {Paunzen}, {H{\"u}mmerich}, \&
  {Bernhard}}]{netopil17}
{Netopil}, M., {Paunzen}, E., {H{\"u}mmerich}, S., \& {Bernhard}, K. 2017,
  \mnras, 468, 2745

\bibitem[{{Perruchot} {et~al.}(2011){Perruchot}, {Bouchy}, {Chazelas},
  {D{\'{\i}}az}, {H{\'e}brard}, {Arnaud}, {Arnold}, {Avila}, {Delfosse},
  {Boisse}, {Moreaux}, {Pepe}, {Richaud}, {Santerne}, {Sottile}, \&
  {T{\'e}zier}}]{perruchot11}
{Perruchot}, S., {Bouchy}, F., {Chazelas}, B., {et~al.} 2011, in \procspie,
  Vol. 8151, Techniques and Instrumentation for Detection of Exoplanets V,
  815115

\bibitem[{{Pickles} \& {Depagne}(2010)}]{pickles10}
{Pickles}, A., \& {Depagne}, {\'E}. 2010, \pasp, 122, 1437

\bibitem[{{Porto de Mello} {et~al.}(2014){Porto de Mello}, {da Silva}, {da
  Silva}, \& {de Nader}}]{mello14}
{Porto de Mello}, G.~F., {da Silva}, R., {da Silva}, L., \& {de Nader}, R.~V.
  2014, \aap, 563, A52

\bibitem[{{Quillen} \& {Thorndike}(2002)}]{quillen02}
{Quillen}, A.~C., \& {Thorndike}, S. 2002, \apjl, 578, L149

\bibitem[{{Renson} \& {Manfroid}(2009)}]{renson09}
{Renson}, P., \& {Manfroid}, J. 2009, \aap, 498, 961

\bibitem[{{Rickman} {et~al.}(2008){Rickman}, {Fouchard}, {Froeschl{\'e}}, \&
  {Valsecchi}}]{rickman08}
{Rickman}, H., {Fouchard}, M., {Froeschl{\'e}}, C., \& {Valsecchi}, G.~B. 2008,
  Celestial Mechanics and Dynamical Astronomy, 102, 111

\bibitem[{{Robinson} {et~al.}(2007){Robinson}, {Ammons}, {Kretke}, {Strader},
  {Wertheimer}, {Fischer}, \& {Laughlin}}]{robinson07}
{Robinson}, S.~E., {Ammons}, S.~M., {Kretke}, K.~A., {et~al.} 2007, \apjs, 169,
  430

\bibitem[{{Roeser} {et~al.}(2010){Roeser}, {Demleitner}, \&
  {Schilbach}}]{roeser10}
{Roeser}, S., {Demleitner}, M., \& {Schilbach}, E. 2010, \aj, 139, 2440

\bibitem[{{Sch{\"o}nrich}(2012)}]{schoenrich12}
{Sch{\"o}nrich}, R. 2012, \mnras, 427, 274

\bibitem[{{Sch{\"o}nrich} {et~al.}(2010){Sch{\"o}nrich}, {Binney}, \&
  {Dehnen}}]{schoenrich10}
{Sch{\"o}nrich}, R., {Binney}, J., \& {Dehnen}, W. 2010, \mnras, 403, 1829

\bibitem[{{Tetzlaff} {et~al.}(2010){Tetzlaff}, {Neuh{\"a}user}, {Hohle}, \&
  {Maciejewski}}]{tetzlaff09}
{Tetzlaff}, N., {Neuh{\"a}user}, R., {Hohle}, M.~M., \& {Maciejewski}, G. 2010,
  \mnras, 402, 2369

\bibitem[{{Thomas} {et~al.}(2016){Thomas}, {Engler}, {Kachelrie{\ss}},
  {Melott}, {Overholt}, \& {Semikoz}}]{thomas16}
{Thomas}, B.~C., {Engler}, E.~E., {Kachelrie{\ss}}, M., {et~al.} 2016, \apjl,
  826, L3

\bibitem[{{van Leeuwen}(2007)}]{leeuwen07}
{van Leeuwen}, F. 2007, \aap, 474, 653

\bibitem[{{Wallner} {et~al.}(2016){Wallner}, {Feige}, {Kinoshita}, {Paul},
  {Fifield}, {Golser}, {Honda}, {Linnemann}, {Matsuzaki}, {Merchel}, {Rugel},
  {Tims}, {Steier}, {Yamagata}, \& {Winkler}}]{wallner16}
{Wallner}, A., {Feige}, J., {Kinoshita}, N., {et~al.} 2016, \nat, 532, 69

\bibitem[{{White} {et~al.}(2007){White}, {Gabor}, \& {Hillenbrand}}]{white07}
{White}, R.~J., {Gabor}, J.~M., \& {Hillenbrand}, L.~A. 2007, \aj, 133, 2524

\bibitem[{{Wickramasinghe}(2010)}]{wickramashinghe10}
{Wickramasinghe}, C. 2010, International Journal of Astrobiology, 9, 119

\bibitem[{{Wielen} {et~al.}(1996){Wielen}, {Fuchs}, \& {Dettbarn}}]{wielen96}
{Wielen}, R., {Fuchs}, B., \& {Dettbarn}, C. 1996, \aap, 314, 438

\bibitem[{{Winters} {et~al.}(2015){Winters}, {Henry}, {Lurie}, {Hambly}, {Jao},
  {Bartlett}, {Boyd}, {Dieterich}, {Finch}, {Hosey}, {Ianna}, {Riedel},
  {Slatten}, \& {Subasavage}}]{winters15}
{Winters}, J.~G., {Henry}, T.~J., {Lurie}, J.~C., {et~al.} 2015, \aj, 149, 5

\bibitem[{{Zacharias} {et~al.}(2017){Zacharias}, {Finch}, \&
  {Frouard}}]{zacharias17}
{Zacharias}, N., {Finch}, C., \& {Frouard}, J. 2017, \aj, 153, 166

\end{thebibliography}
\end{document}